\documentclass[aps,prf,showpacs,floats,twocolumn,floats,superscriptaddress,floatfix,longbibliography]{revtex4-1}
\usepackage{bm}
\usepackage{hyperref}
\usepackage[usenames,dvipsnames]{xcolor}
\usepackage[normalem]{ulem}
\usepackage[utf8]{inputenc}
\usepackage{amsmath}
\usepackage{amsthm}
\usepackage{amsmath}
\usepackage{amsfonts}
\usepackage{blindtext}
\usepackage[utf8]{inputenc}
\usepackage[T1]{fontenc}
\usepackage{graphicx}
\usepackage{float}
\usepackage{wrapfig}
\usepackage{bm}
\usepackage{braket}
\usepackage{grffile}
\usepackage[normalem]{ulem}
\usepackage[usenames,dvipsnames]{xcolor}
\usepackage{mathtools}
\usepackage{xspace}
\usepackage{lineno}
\def\RMV#1{{}}

\newcommand{\be}{\begin{equation}}
\newcommand{\ee}{\end{equation}}

\newcommand{\bu}{\boldsymbol{u}}

\newcommand{\CV}{{\cal V}}

\newcommand{\bX}{\boldsymbol{X}}
\newcommand{\bV}{\boldsymbol{V}}

\newcommand{\kl}[2]{D_{\mathrm{KL}}\!\left(#1 ~ \| ~ #2\right)}

\begin{document}

\title{Synthetic Lagrangian Turbulence by Generative Diffusion Models }

\affiliation{Dept. of Physics and INFN, University of Rome Tor Vergata, Italy.}
\affiliation{Dept. of Industrial Engineering, University of Rome Tor Vergata, Italy.}

\author{T. Li$^{1}$, L. Biferale$^{1}$, F. Bonaccorso$^{1}$,   M. A. Scarpolini$^{2}$, and M. Buzzicotti$^{1}$}
\email{michele.buzzicotti@roma2.infn.it}

\date{\today}
\begin{abstract}
Lagrangian turbulence lies at the core of numerous applied and fundamental problems related to the physics of dispersion and mixing in engineering, bio-fluids, atmosphere, oceans, and astrophysics. Despite exceptional theoretical, numerical, and experimental efforts conducted over the past thirty years, no existing models are capable of faithfully reproducing statistical and topological properties exhibited by particle trajectories in turbulence. We propose a machine learning approach, based on a state-of-the-art diffusion model, to generate single-particle trajectories in three-dimensional turbulence at high Reynolds numbers, thereby bypassing the need for direct numerical simulations or experiments to obtain reliable Lagrangian data. Our model demonstrates the ability to reproduce most statistical benchmarks across time scales, including the fat-tail distribution for velocity increments, the anomalous power law, and the increased intermittency around the dissipative scale. Slight deviations are observed below the dissipative scale, particularly in the acceleration and flatness statistics. Surprisingly, the model exhibits strong generalizability for extreme events, producing events of higher intensity and rarity that still match the realistic statistics. This paves the way for producing synthetic high-quality datasets for pre-training various downstream applications of Lagrangian turbulence.
\end {abstract}

\maketitle

\begin{figure*}
\includegraphics[width=1.0\linewidth]{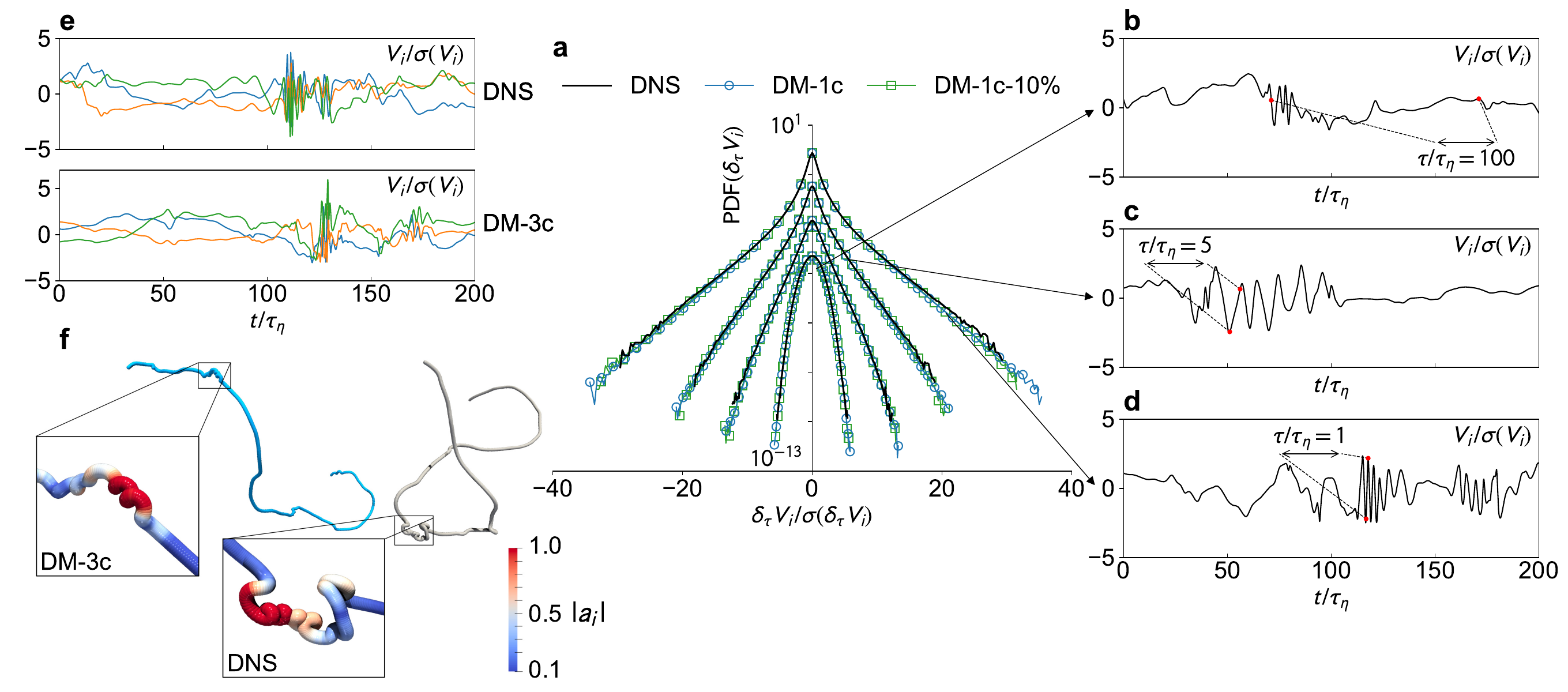}
\caption{\textbf{Comparison between direct numerical simulations (DNS) and diffusion models (DMs).} \textbf{a}, Standardized probability density functions (PDFs) of one generic component of the velocity increment, $\delta_\tau V_i$, at $\tau/\tau_\eta=1,2,5,100$ for ground-truth DNS data (black lines), synthetically generated data from DM-1c (blue lines with circles) and that from DM-1c-10\% (green lines with squares), a DM-1c model trained with 10\% DNS data. PDFs for different $\tau$ are vertically shifted for the sake of presentation. \textbf{b,c,d}, DM-1c trajectories for one generic velocity component with large, medium, and small time increments, $\tau/\tau_\eta=100,5,1$, respectively. \textbf{e}, Comparison of 3D trajectories showing small-scale vortex structures, for both DNS and DM-3c data, where different curves correspond to the three standardized velocity components, $i=x,y,z$. For the DNS, the high oscillatory correlations between the three components are consistent with the presence of strong vortical structures. Similarly, in the case of DM-3c, these correlations can be interpreted as reflecting vortical structures within the hypothetical Eulerian flow. \textbf{f}, Examples of 3D trajectories reconstructed from DNS (bottom) and DM-3c (top). Notice in panel \textbf{a} the remarkable generalizability properties of our DM data-driven model, able to explore and capture extreme events for velocity fluctuations with far larger intensities than observed in the DNS dataset, represented by much more extended tails, while still maintaining the ground truth statistics inherent in the training data. Here, the statistics for DM-1c and DM-1c-10\% data are derived from 86 and 22 times the number of trajectories in the DNS, respectively.}
\label{figure:Figure1}
\end{figure*}

\begin{figure}[h]
\includegraphics[width=1.0\linewidth]{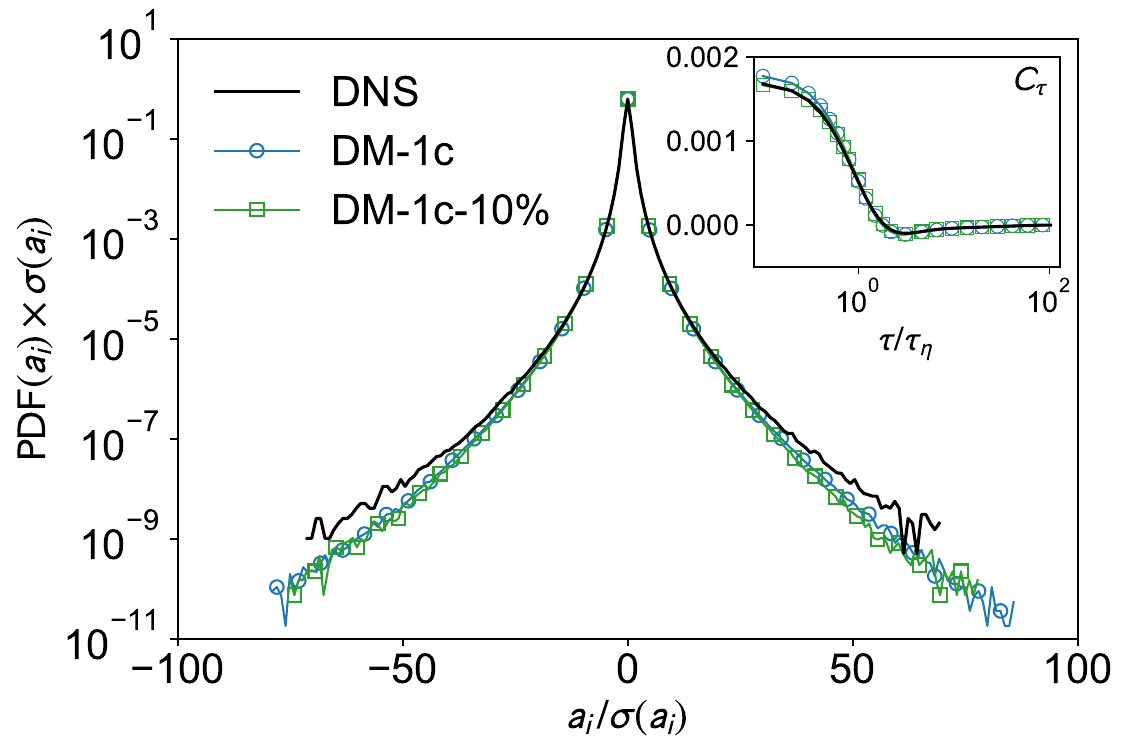}
\caption{\textbf{Statistics of acceleration.} Standardized PDFs of one generic component of the acceleration, $a_i$, for ground-truth DNS data (black line), synthetically generated data from DM-1c (blue line with circles) and that from DM-1c-10\% (green line with squares). Notice the ability of DM-1c to well generalize the statistical trend for rare intense fluctuations never experienced during the training phase with the DNS data. The statistics of the DM-1c and DM-1c-10\% data are based on 86 and 22 times the number of trajectories in the DNS, respectively. Inset: acceleration correlation function.}
\label{figure:Figure2}
\end{figure}

Understanding  the statistical and geometrical properties of particles advected by turbulent flows is a challenging problem of utmost importance for modeling, predicting, and controlling many applications such as combustion, industrial mixing, pollutant dispersion, quantum fluids, protoplanetary disks accretion, cloud formation, prey-predator dynamics, to cite just a few \cite{boris2000scalar, la2001fluid, mordant2001measurement, falkovich2001particles, yeung2002lagrangian, pomeau2016long, falkovich2006lessons, toschi2009lagrangian, shaw2003particle, mckee2021turbulence, bentkamp2019persistent, sawford2013lagrangian, xia2013lagrangian, barenghi2014introduction, xu2014flight, laussy2023shining}. The main difficulties arise from the vast range of time scales involved, spanning from the longest, $\tau_L$, governed by the stirring mechanism to the shortest, $\tau_\eta$, associated with viscous dissipation, and the presence of strong non-Gaussian fluctuations (intermittency). Indeed, the ratio $\tau_L/\tau_\eta$ is proportional to the Taylor Reynolds number, $R_\lambda$, a dimensionless measure of the turbulent intensity, varying from a few thousand in laboratory experiments to millions and even larger in atmospheric and astrophysical contexts \cite{frisch1995turbulence}. Similarly, non-Gaussian fat tails become more pronounced with increasing $R_\lambda$, resulting in rare-but-intense velocity and acceleration fluctuations of up to 50-60 standard deviations that can be easily measured even in table-top laboratory flows at moderate $R_\lambda$ (see Fig. \ref{figure:Figure1}\textbf{a} and Fig. \ref{figure:Figure2}). Due to the combined influence of long-distance sweeping, multi-time fluctuations, and small-scale trapping within intense mini-tornadoes, the problem remains insurmountable from both theoretical and modeling perspectives at the present time.

Over the past thirty years, many different Lagrangian phenomenological models have been proposed, employing various methods such as two-time Ornstein-Uhlembeck stochastic approaches, to capture the dynamics at the two spectrum extremes, $\tau_L, \tau_\eta$ \cite{forcingsawford, pope2011simple}, as well as multi-time infinite-differentiable processes \cite{viggiano2020modelling}. Numerous other models have explored with differing degrees of success, including applications to passive scalar fluctuations \cite{lamorgese2007conditionally, minier2014guidelines, wilson1996review, bourlioux2006conditional, majda2013elementary}. Moreover, both Markovian and non-Markovian modelization based on multifractal and/or multiplicative models, have been employed previously to reproduce certain observed Lagrangian and Eulerian multiscale turbulent features \cite{biferale1998mimicking,arneodo1998random,bacry2003log,chevillard2019skewed,sinhuber2021multi,lubke2022stochastic}, see \cite{zamansky2022acceleration} for a recent attempt to combine multifractal scaling and stochastic partial differential equations. However, although all these previous attempts are able to reproduce well some non-trivial features of the turbulent statistics, we still lack a systematic way to generate synthetic trajectories with the correct multiscale statistics over the full range of dynamics encountered in a real turbulent environment, from the large forcing scales, through the intermittent inertial range, to the coupled regime between inertial and dissipative scales \cite{arneodo2008universal}.

As a result, new approaches are needed to attack the problem. Machine learning (ML) synthetic data-driven models, including variational autoencoders (VAEs) \cite{kingma2014auto}, generative adversarial networks (GANs) \cite{goodfellow2014generative}, and more recently, diffusion models (DMs) \cite{ho2020denoising}, have exhibited remarkable success across diverse fields such as computer vision, audio generation, natural language processing, healthcare, and various other domains \cite{dhariwal2021diffusion,oord2016wavenet,brown2020language,chen2021synthetic}. Building upon this success, there is a growing interest in applying these techniques to scientific challenges. Specifically, ML methods have shown strong potential to tackle open problems in fluid mechanics~\cite{duraisamy2019turbulence,brunton2020machine}. ML tools have been further developed for tasks like generation, super-resolution, prediction, and inpainting of dynamical systems \cite{vlachas2018data, pathak2018model}, two-dimensional (2D) and three-dimensional (3D) Eulerian turbulent snapshots \cite{mohan2020spatio, kim2020deep, guastoni2021convolutional, buzzicotti2021reconstruction, yousif2023deep, shu2023physics}, see \cite{buzzicotti2023data} for a short summary. In many cases, the validation of these tools when applied to fluid mechanics is primarily limited to simple 2D smooth and quasi-Gaussian turbulent flows, or focused on single-point measurements such as mean profiles and two-point spectral properties. There is often a lack of comprehensive quantitative assessments concerning the more intricate multiscale non-Gaussian properties at high Reynolds numbers. Recently, a fully convolutional model has been proposed to generate one-dimensional Eulerian cuts of high-Reynolds-number turbulence \cite{granero2024neural}. This model has demonstrated success in capturing up to the 4th-order structure function, however, its generalization to higher-order statistics exhibits less accuracy. Given the state-of-the-art, it is fair to say that we lack both equation-informed and data-driven tools to generate 3D single- or multi-particle Lagrangian trajectories possessing statistical and geometrical properties that quantitatively agree with experiments and direct numerical simulations (DNS). The demand for the synthetic generation of high-quality and high-quantity data is crucial in various turbulent applications, particularly in the Lagrangian domain, where having even a single trajectory requires the reproduction of the entire Eulerian field over huge spatial domains, which is often a daunting or impossible task for DNS or extremely laborious for experiments.

Here we present a stochastic data-driven model able to match numerical and experimental data concerning single-particle statistics in homogeneous and isotropic turbulence (HIT) at high Reynolds numbers. The model is based on a novel application of state-of-the-art generative DM \cite{ho2020denoising, nichol2021improved, dhariwal2021diffusion}. We have trained two distinct DMs for our study: DM-1c, which generates a single component of the Lagrangian velocity, and DM-3c, which simultaneously outputs all three correlated components (see Methods). Our synthetic generation protocol is able to reproduce the scaling of velocity increments over the full range of available frequencies and for all statistically converged moments up to the 8th order in the original training data. Moreover, the protocol successfully captures acceleration fluctuations of up to 60 standard deviations and even beyond, including the cross-correlations between the three velocity components. We train the model using high-quality data obtained from DNS at $R_\lambda \simeq 310$. The results show also excellent agreement with the numerical ground-truth data for the generalized flatness of 4th, 6th, and 8th orders, whose intensities, due to the presence of intermittent fluctuations, are found to be order of magnitude larger than the expected values in the presence of a Gaussian statistic. Remarkably, our model exhibits strong generalization properties, enabling the synthesis of events with intensities never encountered during the training phase. These extreme fluctuations, resulting from small-scale vortex trapping and sharp u-turn trajectories with unprecedented excursions and rarity, consistently follow the realistic statistics inherent in the training data.

\section{Problem set-up}
\noindent {\bf Lagrangian turbulence.}
The dataset used for training is extracted from a high-resolution DNS of the 3D Navier-Stokes equations (NSE) in a cubic periodic domain with large-scale isotropic forcing. Lagrangian point-like particles have an instantaneous velocity, $\bV(t)=\dot{\bX}(t)$, coinciding with the local instantaneous flow streamlines at the particle position, $\bX(t)$:
\begin{equation}
    \dot{\bX}(t)=\bu(\bX(t),t),
\end{equation}
where $\bu$ solves the NSE, see Eq.~\eqref{eq:nse} in Methods. To construct a high-quality ground-truth database, we tracked a total of ${N}_{p} =327680$ trajectories, each spanning a length of $T\simeq1.3\tau_L\simeq200\tau_\eta$, with a temporal sampling interval of $dt_{s}\simeq 0.1\tau_\eta$. Consequently, each trajectory is discretized into a total of $K=2000$ points, see Table \ref{table:Table1}. Particles are injected randomly in the 3D volume once a statistically stationary evolution is reached for the underlying Eulerian flow, thus ensuring that the Lagrangian statistics are also stationary. The set of multi-time observables utilized to benchmark the quality of the single-particle 3D trajectory generation primarily relies on the statistics of Lagrangian velocity increments:
\begin{equation}
    \delta_\tau V_i(t)=V_i(t+\tau)-V_i(t),
\end{equation}
where $i=x,y,z$ indicates any of the three velocity components and $\tau$ represents the time increment. The instantaneous particle acceleration is obtained from the limit $a_i(t)=\lim_{\tau\to0}\delta_\tau V_i/\tau$, where we use a time resolution of $0.1\tau_{\eta}$ for both DNS and DM. Both the probability density functions (PDFs) of $\delta_\tau V_i$ in Fig.~\ref{figure:Figure1}\textbf{a}, and that of $a_i$ in Fig. \ref{figure:Figure2}, show strongly non-Gaussian fluctuations. The PDFs of $\delta_\tau V_i$ become more pronounced at decreasing the time scale $\tau$. It is a well-known empirical fact that Lagrangian velocity increments develop scaling power-laws in the inertial range, $\tau_\eta\ll\tau\ll\tau_L$, as measured by the Lagrangian Structure Functions \cite{chevillard2003lagrangian,biferale2004multifractal,arneodo2008universal}: 
\begin{equation}
    S_\tau^{(p)}=\langle(\delta_\tau V_i)^p\rangle\sim\tau^{\xi(p)},
\label{eq:SF}
\end{equation}
where with $\langle\cdot\rangle$ we indicate an average over all ${N}_{p}$ trajectories and over time. For both DNS and DM-3c, $S_\tau^{(p)}$ is calculated by further averaging over all velocity components. Henceforth, we neglect the dependence on the velocity component because of isotropy. Concerning the scaling exponents, ${\xi}(p)$, there exists a whole spectrum of anomalous corrections, $\Delta(p)$, to the mean-field dimensional estimate, $p/2$, leading to ${\xi}(p)=p/2+\Delta(p)$. Furthermore, beyond global scaling laws, the statistics of velocity ﬂuctuations can be quantitatively captured scale-by-scale, for each $\tau$ by measuring the local scaling exponents, which are obtained from the logarithmic derivatives of $S_\tau^{(p)}$:
\begin{equation}
    \zeta(p,\tau)=\frac{d\,\log S_\tau^{(p)}}{d\,\log S_\tau^{(2)}}.
\label{eq:chi}
\end{equation}

\begin{table}[h!]
\centering
\begin{tabular}{|c|c|c|c|}
\hline
$N_L$ & $L$    & $dt$               & $\nu$ \\  
1024  & $2\pi$ & $1.5\times10^{-4}$ & $8\times10^{-4}$ \\ \hline
$\epsilon$  & $\tau_\eta$                & $\eta$                     & $R_\lambda$ \\
$1.8\pm0.1$ & $(2.1\pm0.2)\times10^{-2}$ & $(4.2\pm0.1)\times10^{-3}$ & $\simeq 310$ \\ \hline
${ N}_{p}$ & $dt_{s}$            & $T$   & $K$ \\
$327680$   & $2.25\times10^{-3}$ & $4.5$ & $2000$ \\ \hline
\end{tabular}
\caption{\textbf{Eulerian and Lagrangian DNS parameters.}  $N_L$ is the resolution in each spatial dimension; $L$ is the physical dimension of the cubic periodic box; $dt$ represents the time step in the DNS integration; $\nu$ stands for kinematic viscosity; $\epsilon=\nu\langle\partial_i u_j\partial_i u_j\rangle$ is the total mean energy dissipation, averaged over time and space; $\tau_\eta=\sqrt{\nu/\epsilon}$ is the Kolmogorov dissipative time; $\eta=(\nu^3/\epsilon)^{1/4}$ is the Kolmogorov dissipative scale; $R_\lambda=u_{rms}\lambda/\nu$ signifies the `Taylor-scale' Reynolds number, where $u_{rms}$ is the root mean squared velocity, and $\lambda=\sqrt{5E_{tot}/\Omega}\simeq0.14$ represents the `Taylor-scale', with $E_{tot}\simeq4.5$ and $\Omega\simeq1200$, being respectively the total mean energy and enstrophy in the flow. Additionally, $\tau_L=L/u_{rms}\simeq 3.5$ is the integral time scale. Parameters of the Lagrangian particles: $N_{p}$ is the total number of trajectories; $dt_{s}$ is the time lag between two consecutive Lagrangian dumps; $T$ is the total length of each trajectory; and $K=T/dt_{s}$ is the total number of points in each trajectory.}
\label{table:Table1}
\end{table}

\begin{figure*}
\includegraphics[width=1.0\linewidth]{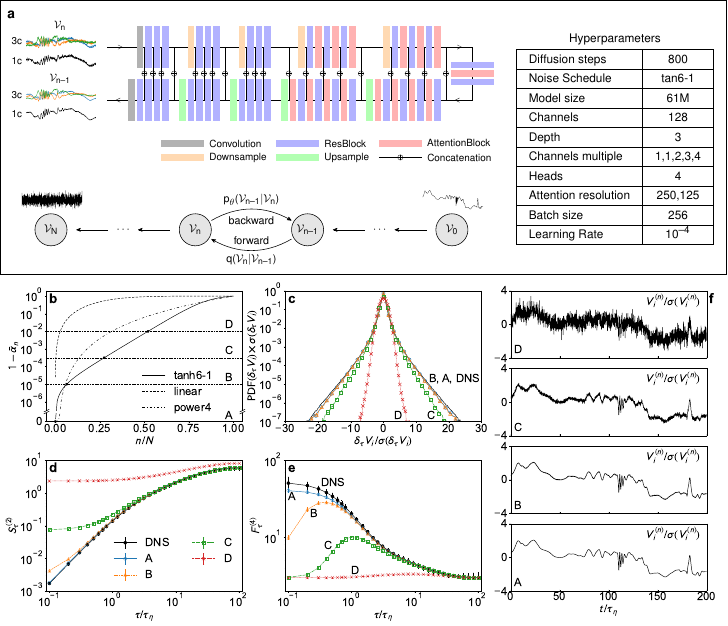}
\caption{\textbf{Illustration of the DM model and in-depth examination of its backward generation process.} \textbf{a}, Schematic representation of the DM model and associated UNet sketch, complemented by a table of hyperparameters. Here, $N$ denotes the total number of diffusion steps and $n$ denotes the intermediate step. More details on the network architecture can be found in the Methods section and in \cite{dhariwal2021diffusion}. \textbf{b}, Three distinct noise schedules for the DM's forward and backward processes explored in this study (see Methods). Points A-D indicate four different stages during the backward generation process (from $\CV_N$ to $\CV_0$) along the optimal noise schedule, curve (tanh6-1). At an early step during the backward process, we have very noisy signals, $n=0.52N$ (D), followed by two intermediate steps at $n=0.27N$ (C) and $n=0.06N$ (B), and the final synthetic trajectory obtained for $n=0$ (A). Please see panel \textbf{f} for the corresponding illustration of one trajectory generation from D to A. A few statistical properties of the DM-1c signals generated at the four backward steps A-D: \textbf{c}, PDF of $\delta_\tau V_i$ for $\tau=\tau_\eta$; \textbf{d}, Second-order structure function, $S^{(2)}_\tau$; \textbf{e}, Fourth-order flatness, $F^{(4)}_\tau$.}
\label{figure:Figure3}
\end{figure*}

\vspace{10pt}
\noindent {\bf Diffusion Models.}
DMs emerge in recent years, outperforming the current state-of-the-art GANs on image synthesis \cite{dhariwal2021diffusion}. DMs are built upon forward and backward diffusion processes (see Fig. \ref{figure:Figure3}\textbf{a} and Methods). The forward process is a Markov chain that gradually introduces Gaussian noise into the training data until the original signal is transformed into pure noise. In the opposite direction, the backward process starts from pure Gaussian-noise realizations and learns to progressively denoise the signal, effectively generating the desired data samples, as shown in Fig. \ref{figure:Figure3}\textbf{f}. The diffusion processes stem from non-equilibrium statistical physics, leveraging Markov chains to progressively morph one distribution into another \cite{sohl2015deep, burda2015accurate}. The training of DMs involves the use of variational inference lower bound to estimate the loss function along a finite, but large, number of diffusion steps. By focusing on these small incremental changes, the loss term becomes tractable, eliminating the need to resort to the less stable adversarial training, a strategy commonly used by GAN, which aims to reproduce the entire data distribution in a single jump from the input noise.
Our implementation of DM has adopted the UNet architecture of the cutting-edge DM model used in computer vision \cite{dhariwal2021diffusion}. An optimized noise schedule for the diffusion processes has also been developed in order to enhance both the efficiency and performance when constructing the multiscale features of the signal, as presented in Fig.~\ref{figure:Figure3}\textbf{b} and discussed in more detail in the Methods section.

\section{Results}

\noindent {\bf Probability density functions.} In Fig. \ref{figure:Figure1}\textbf{a} we show the success of the DM to generate more and more intense (non-Gaussian) velocity fluctuations, $\delta_\tau V_i$, by sending $\tau\to0$, with very good statistical agreement with the ground truth. The typical trajectories generated by the DM-1c are also qualitatively shown in Fig. \ref{figure:Figure1}\textbf{b--d} for different time lags, $\tau$, with local events belonging to both laminar and intense fluctuations. Note the ability of DMs to overcome the additional difficulty of simultaneously generating the three correlated components (DM-3c), required to produce highly complex topological -vortical- structures, as show in Fig. \ref{figure:Figure1}\textbf{e,f}. In Fig. \ref{figure:Figure2} we present the PDF of one generic component of the acceleration, $a_i$, from DM-1c, showing a very close agreement with the fat-tail ground-truth DNS distribution up to fluctuations around 60-70 times the standard deviation. To illustrate the convergence and generalizability of the DM models, we included results in Fig. \ref{figure:Figure1}\textbf{a} and Fig. \ref{figure:Figure2} from the DM-1c model trained on only 10\% of the DNS data, denoted as DM-1c-10\%. The DM-1c and DM-1c-10\% results closely match, demonstrating the training convergence. In Fig. \ref{figure:Figure1}\textbf{a}, the alignment of DM-1c-10\% with the DNS data further underscores the DM's generalizability to generate extreme events unseen in the training data, which importantly, adhere to the realistic statistical properties. Further details and comparisons of other statistical measurements for DM-1c-10\% are provided in the Supplementary Material.

\begin{figure*}
\includegraphics[width=1.0\linewidth]{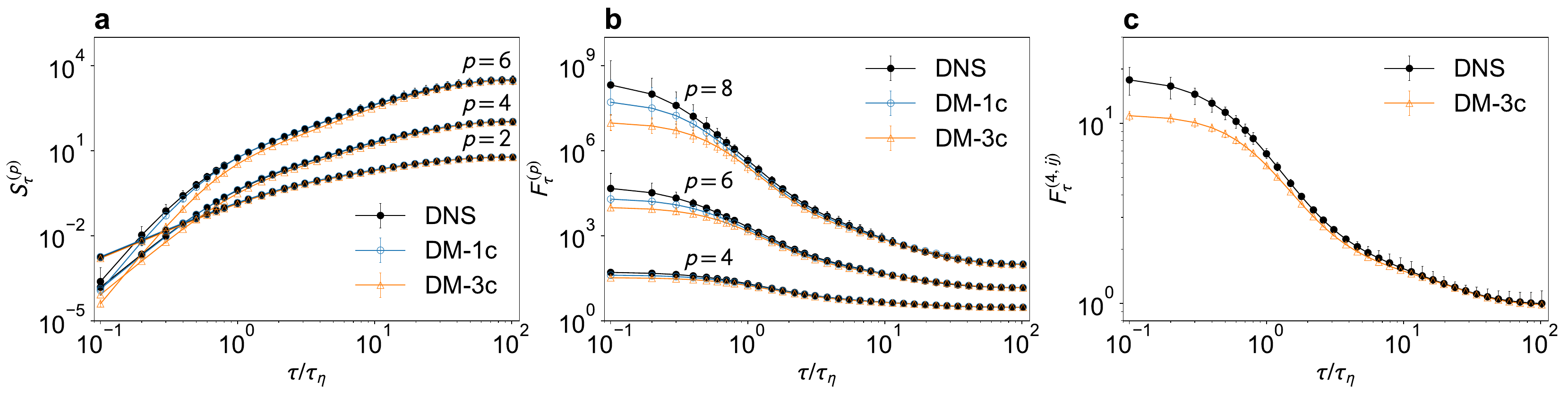}
\caption{\textbf{Multiscale statistical properties of velocity increments.} \textbf{a}, Log-log plot of Lagrangian structure functions, $S^{(p)}_\tau$, for $p=2,4$ and $6$, compared across DNS, DM-1c, and DM-3c. \textbf{b}, Log-log plot of the generalized  flatness, $F^{(p)}_\tau$, for $p=4,6$ and $8$, compared across DNS, DM-1c, and DM-3c. \textbf{c}, Log-log plot of $4$th-order mixed flatness, $F^{(4,ij)}_\tau$, averaged over combinations of $ij=xy,xz$ and $yz$ for both DNS and DM-3c. Error bars are computed as min-max range over the fluctuations of 10 different independent batches sub-sampled from $N_p$ trajectories for each velocity component. Error bars may appear smaller than the data points.}
\label{figure:Figure4}
\end{figure*}

\vspace{10pt}
\noindent {\bf Lagrangian Structure Functions and Generalized Flatness.} In Fig. \ref{figure:Figure4} we show for both DM-1c and DM-3c the Lagrangian structure functions given by (\ref{eq:SF}) for $p=2,4,6$ in panel \textbf{a}, and in panel \textbf{b} the generalized flatness,
\begin{equation}
F^{(p)}_\tau=S^{(p)}_\tau/[S^{(2)}_\tau]^{p/2}.
\end{equation}
Due to the zero-value odd-order structure functions caused by the symmetry of PDFs of the velocity increments, we focus only on the even orders. Structure functions and generalized flatness of different orders are superimposed with the ground-truth DNS for comparison. The capacity of both DM-1c and DM-3c to reproduce the ground truth over many time-scale decades is striking, especially for $\tau\gtrsim\tau_\eta$. However, under the dissipative scale, with $\tau\to0$ we observe a tendency for the DM-3c model to generate a slightly smoother signal compared to the DNS, consistent with our observations in Fig. \ref{figure:Figure2}. The 4th-order mixed flatness, $F^{(4,ij)}_\tau=\langle(\delta_\tau V_i)^2(\delta_\tau V_j)^2\rangle/[S^{(2)}_\tau]^2$, calculated by averaging over $ij=xy,xz$ and $yz$, is shown in panel \textbf{c} of the same figure, in order to check the ability of the DM-3c to reproduce the correlation among different components of the velocity vector, confirming quantitatively the agreement between DM-3c and DNS shown in Fig. \ref{figure:Figure1}\textbf{e,f}. It is worth noting that while the results are very good, there is still room for further refinement of the scales in the dissipative range.

\vspace{10pt}
\noindent {\bf Acceleration correlation function.} In the inset of Fig. \ref{figure:Figure2} we also present the synthetic single-component acceleration correlation function, $C_\tau=\langle a_i(t+\tau)a_i(t)\rangle$, where $i=x,y,z$. The result demonstrates a strong alignment with the DNS. This multiscale Lagrangian structure function has been the subject of intense studying and modeling in the past, due to the presence of a whole set of hierarchical time scales affecting its properties \cite{mordant2002long, angriman2022multitime, mitra2004varieties, l1997temporal}.

\begin{figure*}
\includegraphics[width=1.0\linewidth]{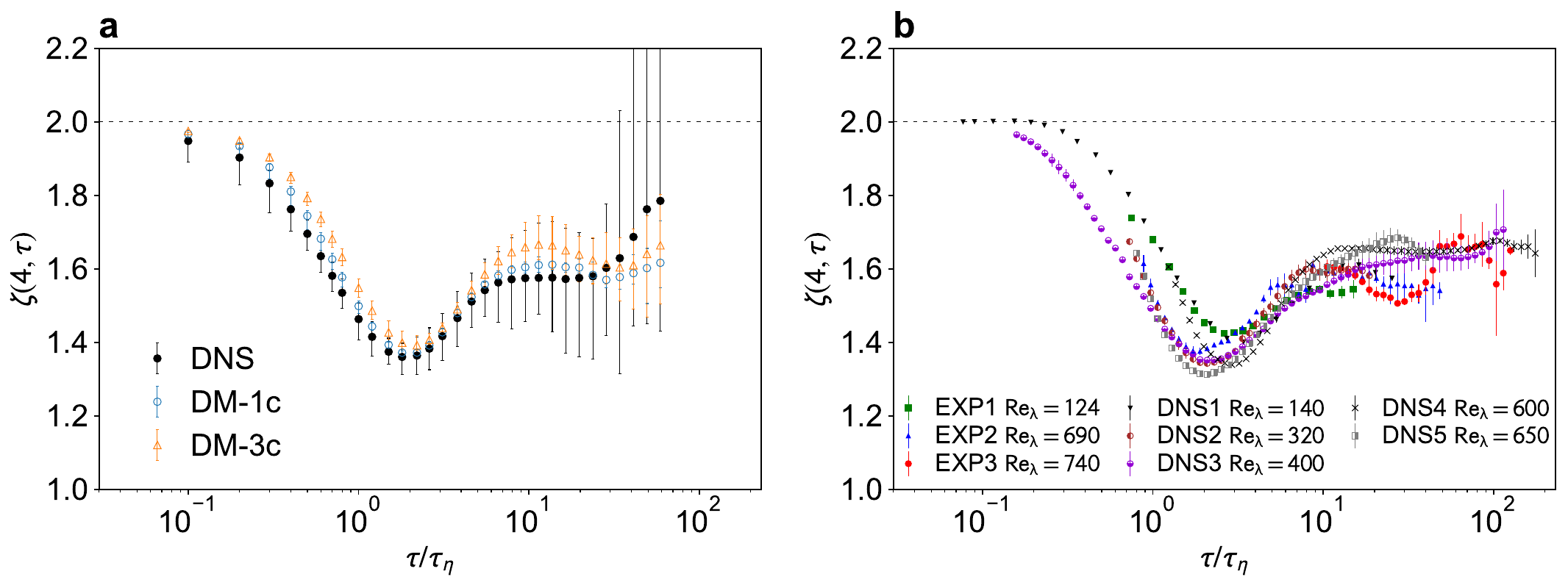}
\caption{\textbf{Scale-by-scale intermittent properties.} \textbf{a}, Comparison between the ground-truth DNS and the two DMs, on the lin-log scale, for the 4th-order logarithmic local slope $\zeta(4,\tau)$ defined in (\ref{eq:chi}). \textbf{b}, The same quantity shown in \textbf{a} from a state-of-the-art collection of DNS \cite{mordant2004experimental,homann2007lagrangian,biferale2005particle,fisher2008terascale,yeung2006reynolds} and experimental data \cite{berg2006backwards,xu2006high,mordant2001measurement} (redrawn from Fig.1 of \cite{arneodo2008universal}). The dotted horizontal lines represent the non-intermittent dimensional scaling, $S^{(4)}_\tau \propto [S^{(2)}_\tau]^2$. Statistics and error bars in \textbf{a} are derived as in Fig. \ref{figure:Figure4}. This resulted in 30 batches for DNS and DM-3c, and 10 batches for DM-1c. The error bars in panel \textbf{b} are computed solely over the three different velocity components.}
\label{figure:Figure5}
\end{figure*}

\vspace{10pt}
\noindent {\bf Local Scaling Exponents.} Let us now introduce what is perhaps the most stringent and quantitative multiscale test for turbulence studies: the comparison of local scaling properties provided by the scale-by-scale exponent defined in (\ref{eq:chi}). In practice, we compute $\zeta(p,\tau)$ by first computing $d\log S_\tau^{(p)}/d\log\tau$ and $d\log S_\tau^{(2)}/d\log\tau$ on a grid with $\tau$ intervals of 1 (from 1 to 1024) using second-order accurate central differences and then performing the division. It is easy to realize that in the inertial range, where (\ref{eq:SF}) is supposed to hold, we have ${\zeta}(p,\tau)={\xi}(p)/{\xi}(2)$, independently of $\tau$. On the other hand, it is known that most of the `turbulent' deadlocks develop at the interface between viscous and inertial ranges, $\tau\sim\tau_\eta$, where the highest level of non-Gaussian fluctuations is observed. Multifractal statistical models are able to fit the whole complexity of the ${\zeta}(p,\tau)$ curves in the entire range of time scales \cite{arneodo2008universal, borgas1993multifractal, chevillard2003lagrangian, nelkin1990multifractal}. This is achieved by introducing a multiplicative cascade model in the inertial range, ended by a {\it fluctuating} dissipative time scale, $\tilde\tau_\eta$ \cite{paladin1987degrees,meneveau1996transition}. Despite numerous attempts, we miss a proper constructive method for embedding the above phenomenology to generate synthetic, realistic 3D Lagrangian trajectories \cite{benzi1993random, arneodo1998random, chevillard2019skewed,zamansky2022acceleration}. In Fig. \ref{figure:Figure5}\textbf{a} we show the local exponent for $p=4$ for DM-1c and DM-3c, and for the DNS data used for training, for comparison in Fig. \ref{figure:Figure5}\textbf{b} we show a state-of-the-art collection of experimental and other DNS data published in the past. Similar results are obtained for $p=6$ and $8$ (not shown). The agreement of results from DMs with experimental and DNS data is remarkable. This is considered a high-quality benchmark, demanding the reproduction of the rate of variation of the {\it local scaling properties} over a range of frequencies/time lags spanning more than $3$ decades and a corresponding variation of the structure functions (\ref{eq:SF}) over 4-5 decades (see Fig. \ref{figure:Figure4}). Such substantial variations are distilled into the measurement of $O(1)$ quantities (\ref{eq:chi}) with an error margin within $5\%$. There are no other tests that can check the scaling properties with greater precision because statistical accuracy typically does not allow one to go beyond a simple - and inaccurate - log-log fit of scaling laws over the full range of variation.

\section{Discussions}

\begin{figure}[h]
\includegraphics[width=1.0\linewidth]{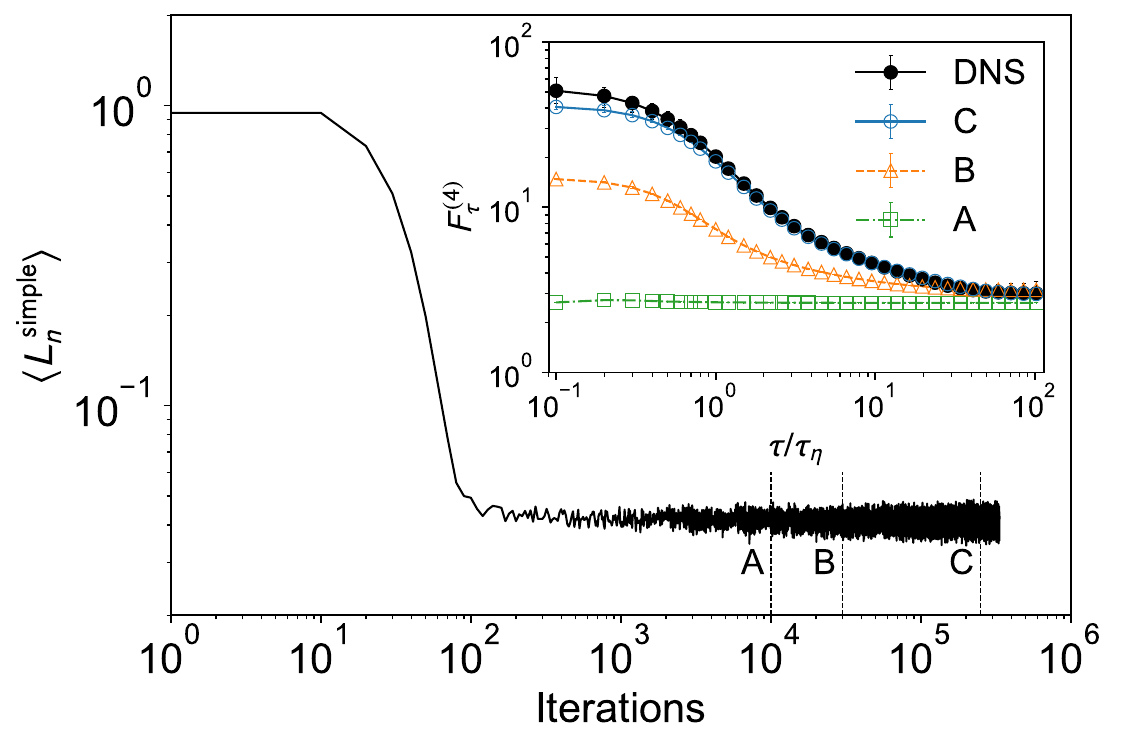}
\caption{\textbf{DM training protocol.} The training loss function, $\langle L_n^\mathrm{simple}\rangle$, against iterations for DM-1c. Here, $\langle\cdot\rangle$ represents the average over a batch of training data, each of which has a corresponding random step $n$ with $0\leq n\leq N$. The inset presents the fourth-order flatness obtained from DM-1c at different iterations (A: $10\times10^3$, B: $30\times10^3$ C: $250\times10^3$), in comparison with that from DNS data. Statistics and error bars are derived as in Fig. \ref{figure:Figure4}.}
\label{figure:Figure6}
\end{figure}

We have presented a data-driven model capable of reproducing all recognized statistical properties of single-particle Lagrangian turbulence in HIT from the large scales down to the inertial and inertial-viscous scaling range, including the enhanced intermittent properties observed around $\tau_\eta$. This achievement is summarized by the PDFs of velocity increments in the inertial range and acceleration (Fig. \ref{figure:Figure1} and Fig. \ref{figure:Figure2}), as well as by the structure functions, the flatness among different components and the local scaling exponents as shown in Figs~\ref{figure:Figure4} and \ref{figure:Figure5}. In Table \ref{table:Table2}, we further summarize a comparison of single-time two-point correlations of velocity and acceleration, showing an excellent matching of DM synthetic data with DNS, except for the case of cross correlation among different acceleration components, $\Sigma^A$, where DM-3c gives a smaller value than DNS. This trend is also reflected in the smoother transition observed in the limit $\tau\to0$ for the single- and mixed-component flatness in Figs. \ref{figure:Figure4}\textbf{b,c}. Furthermore, it is important to highlight the ability of both DM-1c and DM-3c to break the deadlock of viscous intermittency by being able to reproduce the dip structure in the local scaling exponents, as shown in Fig. \ref{figure:Figure5} in the range $\tau\sim\tau_\eta$. Fig. \ref{figure:Figure6} shows how DM generation improves multiscale statistics as training progresses. We also evaluated another prominent generative model, the Wasserstein GAN, for this task. Despite efforts to train and select the best performing model, its accuracy was only satisfactory at large and intermediate scales, and failed considerably at smaller time scales. Further details can be found in the Supplementary Material.

\begin{table}[h!]
\centering
\begin{tabular}{|c|c|c|c|}
\hline
  & DNS & DM-1c & DM-3c \\ \hline
E & $3.0$ & $3.0$ & $2.9$ \\ \hline
A & $1.7\times10^{-3}$ & $1.8\times10^{-3}$ & $1.6\times10^{-3}$ \\ \hline
$\Sigma^V$ & $-0.41$ & $\emptyset$ & $-0.39$ \\ \hline
$\Sigma^A$ & $4.4\times10^{-5}$ & $\emptyset$ & $2.4\times10^{-5}$ \\ \hline
\end{tabular}
\caption{\textbf{Single-time second-order correlations.} Quantities are related to both velocity and acceleration for DNS,  DM-1c and DM-3c: $E=1/3\sum_i\langle V_i^2\rangle$, $A=1/3\sum_i\langle a_i^2\rangle$, $\Sigma^V=1/3\sum_{i,j}\langle V_i^2V_j^2\rangle-\langle V_i^2\rangle\langle V_j^2\rangle$, $\Sigma^A=1/3\sum_{i,j}\langle a_i^2a_j^2\rangle-\langle a_i^2\rangle\langle a_j^2\rangle$, where in the last two expressions the summation is only for $ij=xy,xz$ and $yz$.}
\label{table:Table2}
\end{table}

\vspace{10pt}
\noindent {\bf Generalizability.} Having AI models capable to generate high-quality trajectories can considerably increase the availability of well-validated synthetic data for pre-training physical applications based on Lagrangian single-particle dispersion. Even more surprisingly, our DM model shows the ability to generate trajectories with extremely intense events, thus generalizing beyond the information absorbed during the training phase while still preserving realistic statistical properties. This is clearly illustrated by the striking observation of the extended tails of the PDFs measured from the larger dataset generated by the DM, compared to those measured from the smaller set of training data, as shown in Fig. \ref{figure:Figure1}\textbf{a} and Fig. \ref{figure:Figure2}. Currently, our DM model is not configured to generalize to different flow configurations, such as different boundary conditions, forcing mechanisms or higher Reynolds numbers. Achieving this adaptability may require the use of a conditional diffusion model \cite{dhariwal2021diffusion, nichol2021improved}. By integrating data composed of diverse flows and geometries, such a model could interpolate between different setups and adapt to new conditions, providing a promising avenue for future research.

\vspace{10pt}
\noindent {\bf Explainability.} The fundamental physical model learned by the DM to generate the correct set of multi-time fluctuations remains elusive. DM is based on nested non-linear Gaussian denoising, resembling {\it in spirit} the multiscale build-up of fluctuations used in the creation of multifractal signals and measures. The progressive enrichment of signal properties along the backward diffusion process is displayed in Fig.~\ref{figure:Figure3}\textbf{c--f}. In panel \textbf{e} we show quantitatively the build-up of non-trivial flatness at different stages of the backward process. Similarly, but more qualitatively, panel \textbf{f} shows the emerging non-Gaussian and non-trivial properties within a single trajectory, transitioning from a very noisy signal ($n=0.52N$) to the final step of the backward process ($n=0$). Fig.~\ref{figure:Figure3}\textbf{c--f} illustrates that during the generation process, the model initially generates statistics at larger scales and gradually builds up statistics at smaller scales. Decrypting this multiscale process in terms of precise non-linear mapping could lead to important discoveries in our phenomenological understanding of turbulence. A promising approach to enhance the interpretability of the model is to factorize the data with wavelet decomposition and implement DMs to synthesize the wavelet coefficients, conditioning on the low-frequency ones \cite{guth2022wavelet}.

\vspace{10pt}
\noindent {\bf Impact.} Synthetic stochastic generative models offer remarkable advantages. They (i) provide access to open data without copyright or ethical issues connected to real-data usage, (ii) enable the production of high-quality and high-quantity datasets, which can be used to train other models that require such data as input. The ultimate goal is to provide synthetic datasets that enable new models for downstream applications to reach enhanced accuracy, replacing the necessity for {\it real-data pre-training} with {\it synthetic pre-training}. Our study opens the way for addressing many questions for which the use of real Lagrangian trajectories requires an unfeasible computational or experimental effort. These questions include the relative dispersion problem between two or more particles to study Richardson diffusion \cite{salazar2009two,scatamacchia2012extreme}, shape dynamics \cite{biferale2005multiparticle, xu2011pirouette}, data augmentation of datasets for drifter trajectories in specific oceanic applications \cite{roemmich2019future, essink2022characterizing}, generation and classification of inertial particle trajectories \cite{toschi2009lagrangian} and data inpainting \cite{buzzicotti2021reconstruction}.

\section{Methods}

\vspace{10pt}
\noindent {\bf Navier-Stokes simulations for Lagrangian tracers.}\\
We solve the 3D Navier-Stokes equations:
\begin{equation}
\label{eq:nse}
\begin{cases}
\partial_t \bm{u} + \bm{u} \cdot \nabla \bm{u} = - \nabla p
+ \nu\Delta\bm{u} + \mathbf{F} \\
\nabla \cdot \bm{u} = 0
\end{cases}
\,,
\end{equation}
for an incompressible fluid of viscosity $\nu$ \cite{frisch1995turbulence}. The flow is driven to a non-equilibrium statistically steady state by a homogeneous and isotropic forcing, $\mathbf{F}$, obtained via a second-order Ornstein-Uhlenbeck process \cite{forcingsawford}. For the DNS of the Eulerian field, we used a standard pseudo-spectral solver fully dealiased with the two-thirds rule. Details on the simulation can be found in \cite{TURB-Lagr}. Parameters of the DNS used in this work are given in Table~\ref{table:Table1}. The database of Lagrangian trajectories used in this study is dumped each $dt_{s} = 15 dt \simeq 0.1\tau_\eta$ \cite{calascibetta2023optimal}. Lagrangian integration of tracers is obtained via a B-spline 6th-order interpolation scheme to obtain the fluid velocity at the particle position and with a second-order Adams-Bashforth time-marching scheme \cite{van2012efficiency}.

\vspace{10pt}
\noindent {\bf Diffusion Models.}
The specific implementation of DM utilized in this work is based on the recent research \cite{dhariwal2021diffusion}, which demonstrated extremely good performances of DM even in comparison with GAN for image synthesis. The network architecture, depicted in Fig. \ref{figure:Figure3}, relies on the typical UNet structure \cite{ronneberger2015u}, which is commonly used for image analysis tasks as it is designed to capture both high-level contextual information and precise spatial detail. The UNet consists of two primary components: a contracting and an expanding path. Acting as an encoder, the contracting path progressively reduces the spatial dimension of the input data while increasingly extracting abstract features that contain the global context of the input data. The expanding path acts as a decoder, interpreting the learned features and systematically recovering the spatial resolution to generate the final output (see the later section DM architecture and Noise schedule and Fig. \ref{figure:Figure3} for more details).

\vspace{10pt}
\noindent {\bf Training algorithm.} We train two different classes of DM. One to generate a single component of the Lagrangian velocity field (DM-1c) and one for the three components simultaneously (DM-3c). Let us denote each entire trajectory as $\CV$, where $$\CV=\{V_i(t_k)|t_k\in[0,T];i=x,y,z\};\qquad\text{(DM-1c)}$$ and $$\CV=\{V_x(t_k),V_y(t_k),V_z(t_k)\,|t_k\in[0,T]\};\qquad\text{(DM-3c)}$$ and $k=1,\dots,K$ goes over the total number of discretized sampling times for each trajectory (see Table \ref{table:Table1}). The distribution of the ground truth trajectories obtained from DNS of the NSE is denoted as $q(\CV)$. We introduce a {\it forward} noising process, that starts from the ground truth trajectory, $\CV_0=\CV$, and transforms it, after $N$ steps, to a set of trajectories identical to pure random uncorrelated Gaussian noise. This process generates latent variables $\CV_1,\dots,\CV_N$ by introducing Gaussian noise at step $n$ with a variance $\beta_n\in(0,1)$ according to the following formulation
\begin{equation}
    q(\CV_{1:N}|\CV_0) \coloneqq \prod_{n=1}^{N} q(\CV_n | \CV_{n-1}), \label{eq:joint}
\end{equation}
where we have introduced the shorthand notation $\CV_{1:N}$ to denote the entire chain of the ensemble of noisy trajectories $\CV_{1},\CV_{2},\dots,\CV_{N}$, and each step is defined as
\begin{equation}
    q(\CV_n | \CV_{n-1}) \to \CV_n \sim \mathcal{N}(\sqrt{1-\beta_n} \CV_{n-1}, \beta_n \mathbf{I}).
    \label{eq:singlestep}
\end{equation}
Eq.~(\ref{eq:joint}) is obtained using the Markovian property of the $n$ steps in the forward process. For a large enough $N$ and a suitable sequence of $\beta_n$, the latent vector $\CV_N\sim\mathcal{N}(0,\mathbf{I})$ approximates a delta-correlated Gaussian signal with zero mean and unitary variance. A second remarkable property of the above process, which follows from the Gaussian property of the noise introduced at each step \eqref{eq:singlestep}, is that given $\CV_0$, we can sample $\CV_n$ at any given arbitrary $n$ in a closed form, by defining $\alpha_n \coloneqq 1 - \beta_n$ and $\bar{\alpha}_n \coloneqq \prod_{i=0}^{n} \alpha_i$, as
\begin{equation}
    q(\CV_n|\CV_0) \to \CV_n \sim \mathcal{N}(\sqrt{\bar{\alpha}_n} \CV_0, (1-\bar{\alpha}_n) \mathbf{I}).
    \label{eq:marginal}
\end{equation}
In other words, starting from any ground-truth trajectory, $\CV_0$, we can evaluate its corresponding realization after $n$ steps in the forward process as
\begin{equation}
    \CV_n = \sqrt{\bar{\alpha}_n} \CV_0 + \sqrt{1-\bar{\alpha}_n} \epsilon, \label{eq:jumpnoise}
\end{equation}
where $\epsilon \sim \mathcal{N}(\bm{0},\mathbf{I})$. Now, it is clear that if we can reverse the above process and sample from $p(\CV_{n-1}|\CV_n)$, we will be able to generate new true samples starting from the Gaussian-noise input, $p(\CV_N) = \mathcal{N}(\bm{0},\mathbf{I})$. In general, the backward distribution, $p(\CV_{n-1}|\CV_n)$, is unknown. However, in the limit of continuous diffusion (small $\beta_n$), the reverse process has the identical functional form of the forward process \cite{sohl2015deep}. Since $q(\CV_{n}|\CV_{n-1})$ is a Gaussian distribution, and $\beta_n$ is chosen to be small, then $p(\CV_{n-1}|\CV_n)$ will also be a Gaussian. In this way, the UNet needs to model the mean $\mu_{\theta}(\CV_n, n)$ and standard deviation $\Sigma_{\theta}(\CV_n, n)$ of the transition probabilities for all steps in the backward diffusion process:
\begin{equation}
    p_\theta(\CV_{0:N}) = p(\CV_N) \prod_{n=1}^{N} p_\theta(\CV_{n-1} | \CV_{n}),
\end{equation}
where each reverse step can be written as,
\begin{equation}
\label{eq:nn}
    p_{\theta}(\CV_{n-1}|\CV_n) \to \CV_{n-1} \sim \mathcal{N}(\mu_{\theta}(\CV_n, n), \Sigma_{\theta}(\CV_n, n)).
\end{equation}
During training, the optimization involves minimizing the cross entropy, $L_{CE}$, between the ground truth distribution and the likelihood of the generated data,
\begin{align}
\nonumber
L_{CE} \coloneqq - \mathbb{E}_{q(\CV_{0})} \log &\left ( p_\theta(\CV_0)\right ) = \\
- \mathbb{E}_{q(\CV_{0})}& \log \left ( \int p_\theta(\CV_{0:N}) d\CV_{1:N} \right ).
 \label{eq:cross_entr}
\end{align}
However, integrating over all possible backward paths from $1$ to $N$ and averaging over all ground truth data, $\mathbb{E}_{q(\CV_{0})}[..] = \int [..] q(\CV_{0}) d\CV_0$, to evaluate every network update, is beyond being numerically intractable.\\
A way out is to exploit a variational lower bound $L_{VLB}$, for the cross entropy \cite{sohl2015deep}:
\begin{align}
    & L_{CE} \leq \mathbb{E}_{q(\CV_{0})}\mathbb{E}_{p(\CV_{1:N}|\CV_0)}\bigg[\log\frac{p(\CV_{1:N}|\CV_0)}{p_\theta(\CV_{0:N})}\bigg] \coloneqq L_{VLB}.
    \label{eq:vlb}
\end{align}
To make the above expression computable the expectation value can be split into its independent steps. Consequently, it can be rewritten as a summation of several Kullback-Leibler (KL) divergences, $D_{KL}$, plus one entropy term (see details in Appendix B of~\cite{sohl2015deep}). In this way, $L_{VLB}$ becomes
\begin{align}
    \nonumber
    &L_{VLB} = \mathbb{E}_{q(\CV_0)}\bigg[\underbrace{\kl{p(\CV_N|\CV_0)}{p_{\theta}(\CV_N)}}_{L_N} \\
    &+\sum_{n>1}^N \underbrace{\kl{p(\CV_{n-1}|\CV_n,\CV_0)}{p_\theta(\CV_{n-1}|\CV_n)}}_{L_{n-1}} \nonumber \\
    &\underbrace{-\log p_\theta(\CV_0|\CV_1)}_{L_0}\bigg]. \label{eq:vlb_unrol}
\end{align}
The first term, $L_N$, can be ignored during training, as $p(\CV_N|\CV_0)$ does not depend on the network hyper-parameters, and $p_\theta(\CV_N) = \mathcal{N}(0, \mathbf{I})$ is just the Gaussian distribution. Hence, the network must minimize only the terms, $L_{n}$ with $n<N$, to reproduce the entire backward diffusion process and generate correct data. At this point, the last remarkable property that allows each term of the variational lower bound to be written in a tractable way is that the inverse conditional probability can be calculated analytically when conditioned on a particular realization of the ground-truth data. Using Bayes' theorem, we can write
\begin{equation}
\label{eq:reverse_exact}
    p(\CV_{n-1}|\CV_n,\CV_0) = q(\CV_{n}|\CV_{n-1},\CV_0) \frac{q(\CV_{n-1}|\CV_0)}{q(\CV_{n}|\CV_0)}.
\end{equation}
All probabilities in the right-hand side of Eq.~\eqref{eq:reverse_exact} describe forward steps as defined in Eq.~\eqref{eq:singlestep} and Eq.~\eqref{eq:marginal}. Therefore, Eq.~\eqref{eq:reverse_exact} can be regarded as the product of three Gaussians,
\begin{align}
    \nonumber
    p(\CV_{n-1}|\CV_n,\CV_0) \propto 
    &\exp\left (- \frac{(\CV_{n}-\sqrt{\alpha_n}\CV_{n-1})^2 }{2\beta_n}\right ) \\ \label{eq:reverse_explicit}
    \cdot &\exp\left (- \frac{(\CV_{n-1}-\sqrt{\bar{\alpha}_{n-1}}\CV_0)^2 }{2(1-\bar{\alpha}_{n-1})}\right ) \\ \nonumber
    \cdot &\exp\left ( \frac{(\CV_{n}-\sqrt{\bar{\alpha}_{n}}\CV_0)^2 }{2(1-\bar{\alpha}_{n})}\right ), 
\end{align}
which can be rewritten as
\begin{equation}
\label{eq:reverse}
    p(\CV_{n-1}|\CV_n,\CV_0) \to \CV_{n-1} \sim \mathcal{N}(\tilde{\mu}(\CV_n,\CV_0),\tilde{\beta}_n\mathbf{I}),
\end{equation}
where the mean and the standard deviation of the conditioned reverse probability are, respectively,
\begin{equation}    \tilde{\mu}_n(\CV_n,\CV_0)\coloneqq\frac{\sqrt{\bar{\alpha}_{n-1}}\beta_n}{1-\bar{\alpha}_n}\CV_0 + \frac{\sqrt{\alpha_n}(1-\bar{\alpha}_{n-1})}{1-\bar{\alpha}_n}\CV_n \label{eq:mutilde}
\end{equation}
and
\begin{equation}
    \tilde{\beta}_n\coloneqq\frac{1-\bar{\alpha}_{n-1}}{1-\bar{\alpha}_n}\beta_n.
\end{equation}
All terms denoted by $L_{n-1}$, in the variational lower bound, are $D_{KL}$ between the two Gaussians that depend only on the difference between their mean values and standard deviations. Assuming that the standard deviations of the reverse and forward processes are identical, i.e., $\Sigma_\theta=\beta_n\mathbf{I}$, we only need to model the mean values of the backward Gaussians. Consequently, the KL divergence simplifies to the difference between the two mean values, given in Eq.~\eqref{eq:mutilde} and the output of the UNet mode, $\mu_\theta(\CV_n,n)$, in Eq.~\eqref{eq:nn}. From this simplification, it follows that each loss term becomes 
$$
L_{n-1} = \mathbb{E}_{q(\CV_0)}\bigg[ \frac{1}{2 \beta_n} ||\tilde{\mu}_n(\CV_n,\CV_0)-\mu_\theta(\CV_n,n)||^2 \bigg].
$$
Expressing $\CV_0$ in term of $\CV_n$ by inverting \eqref{eq:jumpnoise} and substituting it in \eqref{eq:mutilde}, the mean value of the reverse conditioned probability can be rewritten as,
\begin{equation}
    \tilde{\mu}(\CV_n, \CV_0) = \frac{1}{\sqrt{\alpha_n}} \left( \CV_n - \frac{\beta_n}{\sqrt{1-\bar{\alpha}_n}} \bm{\epsilon}_{\CV_0,n} \right),
\end{equation}
where the subscripts of the noise term, $\bm{\epsilon}_{\CV_0,n}$, indicate that this is the specific noise realization used to obtain $\CV_n$ from $\CV_0$, as defined in Eq.~\eqref{eq:jumpnoise}. Now since $\CV_n$ is known by the network one may re-parameterize the predicted mean $\mu_\theta(\CV_n,n)$ as:
\begin{equation}
    \mu_{\theta}(\CV_n, n) = \frac{1}{\sqrt{\alpha_n}} \left( \CV_n - \frac{\beta_n}{\sqrt{1-\bar{\alpha}_n}} \bm{\epsilon}_{\theta}(\CV_n, n) \right),
\end{equation}
where $\bm{\epsilon}_\theta$ is a function approximator designed to predict $\bm{\epsilon}_{\CV_0,n}$ from $\CV_n$, leading to the following reformulation of the loss terms,
$$
L_{n-1} = \mathbb{E}_{q(\CV_0),\bm{\epsilon}_{_{\CV_0,n}}}\bigg[ \frac{\beta_n}{2 \alpha_n (1-\bar{\alpha}_n)} ||\bm{\epsilon}_{\CV_0,n} - \bm{\epsilon}_{\theta}(\CV_n, n)||^2 \bigg],
$$
namely in the training $\epsilon_\theta$ predicted from the DM is compared with the one used to build up the $\CV_n$ from $\CV_0$. This formulation leads to faster and more stable training \cite{ho2020denoising}. Moreover, it has been shown \cite{ho2020denoising} that one can obtain good results even by performing the training without learning the variance of the reverse process and introducing a simpler, re-weighted loss function defined as
\begin{equation}\label{eq:loss_simple}
    L^{\mathrm{simple}}_{n-1}= \mathbb{E}_{q(\CV_0),\bm{\epsilon}_{_{\CV_0,n}}}\left[ || \bm{\epsilon}_{\CV_0,n} - \bm{\epsilon}_{\theta}(\CV_n, n) ||^2 \right],
\end{equation}
which is identical to the one we implemented in this work. It is worth noting that due to the Gaussian form of $p_\theta(\CV_0|\CV_1)$, $L_0$ results in the same loss function as depicted in Eq.~\eqref{eq:loss_simple}. Therefore, the optimized loss functions can be expressed as $L_n^{\mathrm{simple}}$, where $n$ ranges from $0$ to $N-1$.

\vspace{10pt}
\noindent {\bf DM architecture and Noise schedule.}
The UNet architecture we have implemented is one of the most advanced networks described in the literature, demonstrating state-of-the-art performance in image generation \cite{dhariwal2021diffusion}. It is capable of extracting the hidden, spatially correlated information that is essential both for image generation and for accomplishing our specific task. The details of the architecture, including the hyperparameters, are summarized in the table in Fig. \ref{figure:Figure3}\textbf{a}. Each encoder and decoder part consists of five levels. Progressing to the next level entails doubling or halving the resolution as one passes through an {\it Upsample} or {\it Downsample} layer, respectively. The {\it Depth} parameter controls the number of {\it ResBlocks} with or without {\it AttentionBlocks} at each level. Within each level, layers share the same number of features, which can be determined using the {\it Channels} and {\it Channels multiple} parameters from the table. Attention mechanisms \cite{vaswani2017attention} allow neural networks to prioritize certain regions or features within the data. In this study, we employed multi-head attention with four heads. {\it AttentionBlocks} were utilized at levels with resolutions of 250 and 125. For the DM-1c model, we utilized $250\times 10^3$ iterations, while $400\times 10^3$ iterations were used for the DM-3c model. In each iteration, we sample a batch of training data and assign a random step index $n$ to each sample, then optimize $L_n^\mathrm{simple}$ across the data batch. Fig. \ref{figure:Figure6} shows the training loss as a function of iteration for DM-1c, alongside the fourth-order flatness of samples generated from it at different iteration checkpoints: A, B, and C. Here, C corresponds to the final model. It reveals that while the loss rapidly reached a `plateau', it is crucial to continue training for the model convergence. This is because $\langle L_n^\mathrm{simple}\rangle$ is an average derived from a data batch where each sample is assigned a random $n$, which does not truly represent the inherent loss $L_{CE}$ described in Eq. \eqref{eq:cross_entr}. While $L_{CE}$ can be approximated as the summed expectation of $L_n^\mathrm{simple}$ across the training dataset for $0<n\le N$, direct evaluation of $L_{CE}$ is impractical. Instead, we rely on examining the statistical properties to measure training progress.\\
Concerning the noise schedule to improve the training and sampling protocols, we explored three different laws and found that the optimal one for our application is given in terms of a tanh-profile, see Fig.~\ref{figure:Figure3}\textbf{b}. Indeed, all results shown in the main text and in panels \textbf{c--e} of the same figure have been obtained by following the schedule (tanh6-1):
\begin{equation}
    \bar{\alpha}_n=\frac{-\tanh{(7n/N-6)}+\tanh{1}}{-\tanh{(-6)}+\tanh{1}},
\end{equation}
which allowed us to use $N=800$ diffusion steps rather than $N=4000$ needed for the linear case where the forward process variances are constantly increasing from $\beta_1=10^{-4}$ to $\beta_N=0.02$. As a result, a five-fold improvement in performance is achieved. We also explored an alternative noise schedule (power4) with a functional form: $\bar{\alpha}_n=1-(n/N)^4$, with $N=800$, which resulted to be slightly inferior to (tanh6-1). Note that applying methods to speed up DM sampling with pre-trained models remains worthy of future exploration \cite{song2021denoising, lu2022dpm}.

\vspace{10pt}
\noindent {\bf Computational cost.}
To illustrate the computational cost of our case, the DNS of the Eulerian field takes about 7.2 hours on 4096 cores. This step is essential even to generate a single Lagrangian trajectory. An additional 64\% of the time is required to track 4 million Lagrangian tracers. All training and sampling of the DM models in our study was performed on 4 NVIDIA A100 GPUs. Training takes approximately 1 hour per 10,000 iterations, resulting in approximately 25 hours for DM-1c and 40 hours for DM-3c. Sampling an equivalent number of 4 million trajectories takes about 200 hours.

\section{Data Availability}

The Lagrangian trajectories used in this study, which include the positions, velocities and accelerations of each particle, are available for download from the open access Smart-TURB portal \url{http://smart-turb.roma2.infn.it}, in the TURB-Lagr repository \cite{TURB-Lagr, calascibetta2023optimal}. It is also possible to download from the same repository a minimum dataset for testing the code and the generated Lagrangian trajectories (velocities over time) used for all analyses in this paper. TURB-Lagr is a newly developed database of 3D turbulent Lagrangian trajectories obtained by DNS of the NSE with homogeneous and isotropic forcing. Details on how to download and read the database are also given in the portal. All data related to this study have also been uploaded to the Open Access Repository \cite{li2024data}.

\section{Code Availability}

The code to train the DM model and generate new trajectories can be found at \url{https://github.com/SmartTURB/diffusion-lagr} \cite{SmartTURB2024DiffusionLagr}. A ready-to-run Code Ocean Capsule with the complete environment is available at \url{https://codeocean.com/capsule/0870187/tree/v1} \cite{Li2024synthetic}.

\section{Acknowledgements}

We acknowledge Lorenzo Basile, Antonio Celani, Massimo Cencini, Sergio Chibbaro, Alessandro Londei and Lionel Mathelin for useful discussion. This work was supported by the European Research Council (ERC) under the European Union’s Horizon 2020 research and innovation programme Smart-TURB (Grant Agreement No. 882340), the MeDiTaTe Project (Grant Agreement No. 859836), the MUR-FARE project R2045J8XAW. 

\section{Author Contributions Statement}

T.L., L.B., and M.B. conceived the work. T.L. and M.B. performed all the numerical simulations and data analysis. All authors contributed to the interpretation of the results. L.B., T.L., M.S., and M.B. wrote the manuscript.   

\section{Competing Interests Statement}

The authors declare no competing interests.

\clearpage

\section{References}


\begin{thebibliography}{10}
\expandafter\ifx\csname url\endcsname\relax
  \def\url#1{\texttt{#1}}\fi
\expandafter\ifx\csname urlprefix\endcsname\relax\def\urlprefix{URL }\fi
\providecommand{\bibinfo}[2]{#2}
\providecommand{\eprint}[2][]{\url{#2}}

\bibitem{boris2000scalar}
\bibinfo{author}{Shraiman, I.~B.} \& \bibinfo{author}{D.~Siggia, D.~E.}
\newblock \bibinfo{title}{Scalar turbulence}.
\newblock \emph{\bibinfo{journal}{Nature}} \textbf{\bibinfo{volume}{405}},
  \bibinfo{pages}{639--646} (\bibinfo{year}{2000}).

\bibitem{la2001fluid}
\bibinfo{author}{La~Porta, A.}, \bibinfo{author}{Voth, G.~A.},
  \bibinfo{author}{Crawford, A.~M.}, \bibinfo{author}{Alexander, J.} \&
  \bibinfo{author}{Bodenschatz, E.}
\newblock \bibinfo{title}{Fluid particle accelerations in fully developed
  turbulence}.
\newblock \emph{\bibinfo{journal}{Nature}} \textbf{\bibinfo{volume}{409}},
  \bibinfo{pages}{1017--1019} (\bibinfo{year}{2001}).

\bibitem{mordant2001measurement}
\bibinfo{author}{Mordant, N.}, \bibinfo{author}{Metz, P.},
  \bibinfo{author}{Michel, O.} \& \bibinfo{author}{Pinton, J.-F.}
\newblock \bibinfo{title}{Measurement of lagrangian velocity in fully developed
  turbulence}.
\newblock \emph{\bibinfo{journal}{Physical Review Letters}}
  \textbf{\bibinfo{volume}{87}}, \bibinfo{pages}{214501}
  (\bibinfo{year}{2001}).

\bibitem{falkovich2001particles}
\bibinfo{author}{Falkovich, G.}, \bibinfo{author}{Gaw\ifmmode~\mbox{\c{e}}\else
  \c{e}\fi{}dzki, K.} \& \bibinfo{author}{Vergassola, M.}
\newblock \bibinfo{title}{Particles and fields in fluid turbulence}.
\newblock \emph{\bibinfo{journal}{Rev. Mod. Phys.}}
  \textbf{\bibinfo{volume}{73}}, \bibinfo{pages}{913--975}
  (\bibinfo{year}{2001}).
\newblock \urlprefix\url{https://link.aps.org/doi/10.1103/RevModPhys.73.913}.

\bibitem{yeung2002lagrangian}
\bibinfo{author}{Yeung, P.}
\newblock \bibinfo{title}{Lagrangian investigations of turbulence}.
\newblock \emph{\bibinfo{journal}{Annual review of fluid mechanics}}
  \textbf{\bibinfo{volume}{34}}, \bibinfo{pages}{115--142}
  (\bibinfo{year}{2002}).

\bibitem{pomeau2016long}
\bibinfo{author}{Pomeau, Y.}
\newblock \bibinfo{title}{The long and winding road}.
\newblock \emph{\bibinfo{journal}{Nature Physics}}
  \textbf{\bibinfo{volume}{12}}, \bibinfo{pages}{198--199}
  (\bibinfo{year}{2016}).

\bibitem{falkovich2006lessons}
\bibinfo{author}{Falkovich, G.} \& \bibinfo{author}{Sreenivasan, K.~R.}
\newblock \bibinfo{title}{Lessons from hydrodynamic turbulence}.
\newblock \emph{\bibinfo{journal}{Physics Today}}
  \textbf{\bibinfo{volume}{59}}, \bibinfo{pages}{43} (\bibinfo{year}{2006}).

\bibitem{toschi2009lagrangian}
\bibinfo{author}{Toschi, F.} \& \bibinfo{author}{Bodenschatz, E.}
\newblock \bibinfo{title}{Lagrangian properties of particles in turbulence}.
\newblock \emph{\bibinfo{journal}{Annual review of fluid mechanics}}
  \textbf{\bibinfo{volume}{41}}, \bibinfo{pages}{375--404}
  (\bibinfo{year}{2009}).

\bibitem{shaw2003particle}
\bibinfo{author}{Shaw, R.~A.}
\newblock \bibinfo{title}{Particle-turbulence interactions in atmospheric
  clouds}.
\newblock \emph{\bibinfo{journal}{Annual Review of Fluid Mechanics}}
  \textbf{\bibinfo{volume}{35}}, \bibinfo{pages}{183--227}
  (\bibinfo{year}{2003}).

\bibitem{mckee2021turbulence}
\bibinfo{author}{McKee, C.~F.} \& \bibinfo{author}{Stone, J.~M.}
\newblock \bibinfo{title}{Turbulence in the heavens}.
\newblock \emph{\bibinfo{journal}{Nature Astronomy}}
  \textbf{\bibinfo{volume}{5}}, \bibinfo{pages}{342--343}
  (\bibinfo{year}{2021}).

\bibitem{bentkamp2019persistent}
\bibinfo{author}{Bentkamp, L.}, \bibinfo{author}{Lalescu, C.~C.} \&
  \bibinfo{author}{Wilczek, M.}
\newblock \bibinfo{title}{Persistent accelerations disentangle lagrangian
  turbulence}.
\newblock \emph{\bibinfo{journal}{Nature Communications}}
  \textbf{\bibinfo{volume}{10}}, \bibinfo{pages}{3550} (\bibinfo{year}{2019}).

\bibitem{sawford2013lagrangian}
\bibinfo{author}{Sawford, B.~L.} \& \bibinfo{author}{Pinton, J.-F.}
\newblock \bibinfo{title}{A lagrangian view of turbulent dispersion and
  mixing}.
\newblock In \emph{\bibinfo{booktitle}{Ten Chapters in Turbulance}},
  \bibinfo{pages}{132--175} (\bibinfo{publisher}{Cambridge University Press},
  \bibinfo{year}{2013}).

\bibitem{xia2013lagrangian}
\bibinfo{author}{Xia, H.}, \bibinfo{author}{Francois, N.},
  \bibinfo{author}{Punzmann, H.} \& \bibinfo{author}{Shats, M.}
\newblock \bibinfo{title}{Lagrangian scale of particle dispersion in
  turbulence}.
\newblock \emph{\bibinfo{journal}{Nature communications}}
  \textbf{\bibinfo{volume}{4}}, \bibinfo{pages}{2013} (\bibinfo{year}{2013}).

\bibitem{barenghi2014introduction}
\bibinfo{author}{Barenghi, C.~F.}, \bibinfo{author}{Skrbek, L.} \&
  \bibinfo{author}{Sreenivasan, K.~R.}
\newblock \bibinfo{title}{Introduction to quantum turbulence}.
\newblock \emph{\bibinfo{journal}{Proceedings of the National Academy of
  Sciences}} \textbf{\bibinfo{volume}{111}}, \bibinfo{pages}{4647--4652}
  (\bibinfo{year}{2014}).

\bibitem{xu2014flight}
\bibinfo{author}{Xu, H.} \emph{et~al.}
\newblock \bibinfo{title}{Flight--crash events in turbulence}.
\newblock \emph{\bibinfo{journal}{Proceedings of the National Academy of
  Sciences}} \textbf{\bibinfo{volume}{111}}, \bibinfo{pages}{7558--7563}
  (\bibinfo{year}{2014}).

\bibitem{laussy2023shining}
\bibinfo{author}{Laussy, F.~P.}
\newblock \bibinfo{title}{Shining light on turbulence}.
\newblock \emph{\bibinfo{journal}{Nature Photonics}}
  \textbf{\bibinfo{volume}{17}}, \bibinfo{pages}{381--382}
  (\bibinfo{year}{2023}).

\bibitem{frisch1995turbulence}
\bibinfo{author}{Frisch, U.}
\newblock \emph{\bibinfo{title}{Turbulence: the legacy of AN Kolmogorov}}
  (\bibinfo{publisher}{Cambridge University Press}, \bibinfo{year}{1995}).

\bibitem{forcingsawford}
\bibinfo{author}{{Sawford}, B.~L.}
\newblock \bibinfo{title}{{Reynolds number effects in Lagrangian stochastic
  models of turbulent dispersion}}.
\newblock \emph{\bibinfo{journal}{Phys. Fluids A: Fluid Dyn.}}
  \textbf{\bibinfo{volume}{3}}, \bibinfo{pages}{1577--1586}
  (\bibinfo{year}{1991}).

\bibitem{pope2011simple}
\bibinfo{author}{Pope, S.~B.}
\newblock \bibinfo{title}{Simple models of turbulent flows}.
\newblock \emph{\bibinfo{journal}{Physics of Fluids}}
  \textbf{\bibinfo{volume}{23}}, \bibinfo{pages}{011301}
  (\bibinfo{year}{2011}).

\bibitem{viggiano2020modelling}
\bibinfo{author}{Viggiano, B.} \emph{et~al.}
\newblock \bibinfo{title}{Modelling lagrangian velocity and acceleration in
  turbulent flows as infinitely differentiable stochastic processes}.
\newblock \emph{\bibinfo{journal}{Journal of Fluid Mechanics}}
  \textbf{\bibinfo{volume}{900}}, \bibinfo{pages}{A27} (\bibinfo{year}{2020}).

\bibitem{lamorgese2007conditionally}
\bibinfo{author}{Lamorgese, A.}, \bibinfo{author}{Pope, S.~B.},
  \bibinfo{author}{Yeung, P.} \& \bibinfo{author}{Sawford, B.~L.}
\newblock \bibinfo{title}{A conditionally cubic-gaussian stochastic lagrangian
  model for acceleration in isotropic turbulence}.
\newblock \emph{\bibinfo{journal}{Journal of Fluid Mechanics}}
  \textbf{\bibinfo{volume}{582}}, \bibinfo{pages}{423--448}
  (\bibinfo{year}{2007}).

\bibitem{minier2014guidelines}
\bibinfo{author}{Minier, J.-P.}, \bibinfo{author}{Chibbaro, S.} \&
  \bibinfo{author}{Pope, S.~B.}
\newblock \bibinfo{title}{Guidelines for the formulation of lagrangian
  stochastic models for particle simulations of single-phase and dispersed
  two-phase turbulent flows}.
\newblock \emph{\bibinfo{journal}{Physics of Fluids}}
  \textbf{\bibinfo{volume}{26}}, \bibinfo{pages}{113303}
  (\bibinfo{year}{2014}).

\bibitem{wilson1996review}
\bibinfo{author}{Wilson, J.~D.} \& \bibinfo{author}{Sawford, B.~L.}
\newblock \bibinfo{title}{Review of lagrangian stochastic models for
  trajectories in the turbulent atmosphere}.
\newblock \emph{\bibinfo{journal}{Boundary-layer meteorology}}
  \textbf{\bibinfo{volume}{78}}, \bibinfo{pages}{191--210}
  (\bibinfo{year}{1996}).

\bibitem{bourlioux2006conditional}
\bibinfo{author}{Bourlioux, A.}, \bibinfo{author}{Majda, A.} \&
  \bibinfo{author}{Volkov, O.}
\newblock \bibinfo{title}{Conditional statistics for a passive scalar with a
  mean gradient and intermittency}.
\newblock \emph{\bibinfo{journal}{Physics of Fluids}}
  \textbf{\bibinfo{volume}{18}} (\bibinfo{year}{2006}).

\bibitem{majda2013elementary}
\bibinfo{author}{Majda, A.~J.} \& \bibinfo{author}{Gershgorin, B.}
\newblock \bibinfo{title}{Elementary models for turbulent diffusion with
  complex physical features: eddy diffusivity, spectrum and intermittency}.
\newblock \emph{\bibinfo{journal}{Philosophical Transactions of the Royal
  Society A: Mathematical, Physical and Engineering Sciences}}
  \textbf{\bibinfo{volume}{371}}, \bibinfo{pages}{20120184}
  (\bibinfo{year}{2013}).

\bibitem{biferale1998mimicking}
\bibinfo{author}{Biferale, L.}, \bibinfo{author}{Boffetta, G.},
  \bibinfo{author}{Celani, A.}, \bibinfo{author}{Crisanti, A.} \&
  \bibinfo{author}{Vulpiani, A.}
\newblock \bibinfo{title}{Mimicking a turbulent signal: Sequential multiaffine
  processes}.
\newblock \emph{\bibinfo{journal}{Physical Review E}}
  \textbf{\bibinfo{volume}{57}}, \bibinfo{pages}{R6261} (\bibinfo{year}{1998}).

\bibitem{arneodo1998random}
\bibinfo{author}{Arneodo, A.}, \bibinfo{author}{Bacry, E.} \&
  \bibinfo{author}{Muzy, J.-F.}
\newblock \bibinfo{title}{Random cascades on wavelet dyadic trees}.
\newblock \emph{\bibinfo{journal}{Journal of Mathematical Physics}}
  \textbf{\bibinfo{volume}{39}}, \bibinfo{pages}{4142--4164}
  (\bibinfo{year}{1998}).

\bibitem{bacry2003log}
\bibinfo{author}{Bacry, E.} \& \bibinfo{author}{Muzy, J.~F.}
\newblock \bibinfo{title}{Log-infinitely divisible multifractal processes}.
\newblock \emph{\bibinfo{journal}{Communications in Mathematical Physics}}
  \textbf{\bibinfo{volume}{236}}, \bibinfo{pages}{449--475}
  (\bibinfo{year}{2003}).

\bibitem{chevillard2019skewed}
\bibinfo{author}{Chevillard, L.}, \bibinfo{author}{Garban, C.},
  \bibinfo{author}{Rhodes, R.} \& \bibinfo{author}{Vargas, V.}
\newblock \bibinfo{title}{On a skewed and multifractal unidimensional random
  field, as a probabilistic representation of kolmogorov’s views on
  turbulence}.
\newblock In \emph{\bibinfo{booktitle}{Annales Henri Poincar{\'e}}},
  vol.~\bibinfo{volume}{20}, \bibinfo{pages}{3693--3741}
  (\bibinfo{organization}{Springer}, \bibinfo{year}{2019}).

\bibitem{sinhuber2021multi}
\bibinfo{author}{Sinhuber, M.}, \bibinfo{author}{Friedrich, J.},
  \bibinfo{author}{Grauer, R.} \& \bibinfo{author}{Wilczek, M.}
\newblock \bibinfo{title}{Multi-level stochastic refinement for complex time
  series and fields: a data-driven approach}.
\newblock \emph{\bibinfo{journal}{New Journal of Physics}}
  \textbf{\bibinfo{volume}{23}}, \bibinfo{pages}{063063}
  (\bibinfo{year}{2021}).

\bibitem{lubke2022stochastic}
\bibinfo{author}{L{\"u}bke, J.}, \bibinfo{author}{Friedrich, J.} \&
  \bibinfo{author}{Grauer, R.}
\newblock \bibinfo{title}{Stochastic interpolation of sparsely sampled time
  series by a superstatistical random process and its synthesis in fourier and
  wavelet space}.
\newblock \emph{\bibinfo{journal}{Journal of Physics: Complexity}}
  (\bibinfo{year}{2022}).

\bibitem{zamansky2022acceleration}
\bibinfo{author}{Zamansky, R.}
\newblock \bibinfo{title}{Acceleration scaling and stochastic dynamics of a
  fluid particle in turbulence}.
\newblock \emph{\bibinfo{journal}{Physical Review Fluids}}
  \textbf{\bibinfo{volume}{7}}, \bibinfo{pages}{084608} (\bibinfo{year}{2022}).

\bibitem{arneodo2008universal}
\bibinfo{author}{Arn{\'e}odo, A.} \emph{et~al.}
\newblock \bibinfo{title}{Universal intermittent properties of particle
  trajectories in highly turbulent flows}.
\newblock \emph{\bibinfo{journal}{Physical Review Letters}}
  \textbf{\bibinfo{volume}{100}}, \bibinfo{pages}{254504}
  (\bibinfo{year}{2008}).

\bibitem{kingma2014auto}
\bibinfo{author}{Kingma, D.~P.} \& \bibinfo{author}{Welling, M.}
\newblock \bibinfo{title}{{Auto-Encoding Variational Bayes}}.
\newblock In \emph{\bibinfo{booktitle}{2nd International Conference on Learning
  Representations, {ICLR} 2014, Banff, AB, Canada, April 14-16, 2014,
  Conference Track Proceedings}} (\bibinfo{year}{2014}).
\newblock \eprint{http://arxiv.org/abs/1312.6114v10}.

\bibitem{goodfellow2014generative}
\bibinfo{author}{Goodfellow, I.} \emph{et~al.}
\newblock \bibinfo{title}{Generative adversarial nets}.
\newblock \emph{\bibinfo{journal}{Advances in neural information processing
  systems}} \textbf{\bibinfo{volume}{27}} (\bibinfo{year}{2014}).

\bibitem{ho2020denoising}
\bibinfo{author}{Ho, J.}, \bibinfo{author}{Jain, A.} \&
  \bibinfo{author}{Abbeel, P.}
\newblock \bibinfo{title}{Denoising diffusion probabilistic models}.
\newblock \emph{\bibinfo{journal}{Advances in Neural Information Processing
  Systems}} \textbf{\bibinfo{volume}{33}}, \bibinfo{pages}{6840--6851}
  (\bibinfo{year}{2020}).

\bibitem{dhariwal2021diffusion}
\bibinfo{author}{Dhariwal, P.} \& \bibinfo{author}{Nichol, A.}
\newblock \bibinfo{title}{Diffusion models beat gans on image synthesis}.
\newblock \emph{\bibinfo{journal}{Advances in Neural Information Processing
  Systems}} \textbf{\bibinfo{volume}{34}}, \bibinfo{pages}{8780--8794}
  (\bibinfo{year}{2021}).

\bibitem{oord2016wavenet}
\bibinfo{author}{{van den Oord}, A.} \emph{et~al.}
\newblock \bibinfo{title}{{WaveNet: A Generative Model for Raw Audio}}.
\newblock In \emph{\bibinfo{booktitle}{Proc. 9th ISCA Workshop on Speech
  Synthesis Workshop (SSW 9)}}, \bibinfo{pages}{125} (\bibinfo{year}{2016}).

\bibitem{brown2020language}
\bibinfo{author}{Brown, T.} \emph{et~al.}
\newblock \bibinfo{title}{Language models are few-shot learners}.
\newblock \emph{\bibinfo{journal}{Advances in neural information processing
  systems}} \textbf{\bibinfo{volume}{33}}, \bibinfo{pages}{1877--1901}
  (\bibinfo{year}{2020}).

\bibitem{chen2021synthetic}
\bibinfo{author}{Chen, R.~J.}, \bibinfo{author}{Lu, M.~Y.},
  \bibinfo{author}{Chen, T.~Y.}, \bibinfo{author}{Williamson, D.~F.} \&
  \bibinfo{author}{Mahmood, F.}
\newblock \bibinfo{title}{Synthetic data in machine learning for medicine and
  healthcare}.
\newblock \emph{\bibinfo{journal}{Nature Biomedical Engineering}}
  \textbf{\bibinfo{volume}{5}}, \bibinfo{pages}{493--497}
  (\bibinfo{year}{2021}).

\bibitem{duraisamy2019turbulence}
\bibinfo{author}{Duraisamy, K.}, \bibinfo{author}{Iaccarino, G.} \&
  \bibinfo{author}{Xiao, H.}
\newblock \bibinfo{title}{Turbulence modeling in the age of data}.
\newblock \emph{\bibinfo{journal}{Annual review of fluid mechanics}}
  \textbf{\bibinfo{volume}{51}}, \bibinfo{pages}{357--377}
  (\bibinfo{year}{2019}).

\bibitem{brunton2020machine}
\bibinfo{author}{Brunton, S.~L.}, \bibinfo{author}{Noack, B.~R.} \&
  \bibinfo{author}{Koumoutsakos, P.}
\newblock \bibinfo{title}{Machine learning for fluid mechanics}.
\newblock \emph{\bibinfo{journal}{Annual review of fluid mechanics}}
  \textbf{\bibinfo{volume}{52}}, \bibinfo{pages}{477--508}
  (\bibinfo{year}{2020}).

\bibitem{vlachas2018data}
\bibinfo{author}{Vlachas, P.~R.}, \bibinfo{author}{Byeon, W.},
  \bibinfo{author}{Wan, Z.~Y.}, \bibinfo{author}{Sapsis, T.~P.} \&
  \bibinfo{author}{Koumoutsakos, P.}
\newblock \bibinfo{title}{Data-driven forecasting of high-dimensional chaotic
  systems with long short-term memory networks}.
\newblock \emph{\bibinfo{journal}{Proceedings of the Royal Society A:
  Mathematical, Physical and Engineering Sciences}}
  \textbf{\bibinfo{volume}{474}}, \bibinfo{pages}{20170844}
  (\bibinfo{year}{2018}).

\bibitem{pathak2018model}
\bibinfo{author}{Pathak, J.}, \bibinfo{author}{Hunt, B.},
  \bibinfo{author}{Girvan, M.}, \bibinfo{author}{Lu, Z.} \&
  \bibinfo{author}{Ott, E.}
\newblock \bibinfo{title}{Model-free prediction of large spatiotemporally
  chaotic systems from data: A reservoir computing approach}.
\newblock \emph{\bibinfo{journal}{Physical review letters}}
  \textbf{\bibinfo{volume}{120}}, \bibinfo{pages}{024102}
  (\bibinfo{year}{2018}).

\bibitem{mohan2020spatio}
\bibinfo{author}{Mohan, A.~T.}, \bibinfo{author}{Tretiak, D.},
  \bibinfo{author}{Chertkov, M.} \& \bibinfo{author}{Livescu, D.}
\newblock \bibinfo{title}{Spatio-temporal deep learning models of 3d turbulence
  with physics informed diagnostics}.
\newblock \emph{\bibinfo{journal}{Journal of Turbulence}}
  \textbf{\bibinfo{volume}{21}}, \bibinfo{pages}{484--524}
  (\bibinfo{year}{2020}).

\bibitem{kim2020deep}
\bibinfo{author}{Kim, J.} \& \bibinfo{author}{Lee, C.}
\newblock \bibinfo{title}{Deep unsupervised learning of turbulence for inflow
  generation at various reynolds numbers}.
\newblock \emph{\bibinfo{journal}{Journal of Computational Physics}}
  \textbf{\bibinfo{volume}{406}}, \bibinfo{pages}{109216}
  (\bibinfo{year}{2020}).

\bibitem{guastoni2021convolutional}
\bibinfo{author}{Guastoni, L.} \emph{et~al.}
\newblock \bibinfo{title}{Convolutional-network models to predict wall-bounded
  turbulence from wall quantities}.
\newblock \emph{\bibinfo{journal}{Journal of Fluid Mechanics}}
  \textbf{\bibinfo{volume}{928}}, \bibinfo{pages}{A27} (\bibinfo{year}{2021}).

\bibitem{buzzicotti2021reconstruction}
\bibinfo{author}{Buzzicotti, M.}, \bibinfo{author}{Bonaccorso, F.},
  \bibinfo{author}{Di~Leoni, P.~C.} \& \bibinfo{author}{Biferale, L.}
\newblock \bibinfo{title}{Reconstruction of turbulent data with deep generative
  models for semantic inpainting from turb-rot database}.
\newblock \emph{\bibinfo{journal}{Physical Review Fluids}}
  \textbf{\bibinfo{volume}{6}}, \bibinfo{pages}{050503} (\bibinfo{year}{2021}).

\bibitem{yousif2023deep}
\bibinfo{author}{Yousif, M.~Z.}, \bibinfo{author}{Yu, L.},
  \bibinfo{author}{Hoyas, S.}, \bibinfo{author}{Vinuesa, R.} \&
  \bibinfo{author}{Lim, H.}
\newblock \bibinfo{title}{A deep-learning approach for reconstructing 3d
  turbulent flows from 2d observation data}.
\newblock \emph{\bibinfo{journal}{Scientific Reports}}
  \textbf{\bibinfo{volume}{13}}, \bibinfo{pages}{2529} (\bibinfo{year}{2023}).

\bibitem{shu2023physics}
\bibinfo{author}{Shu, D.}, \bibinfo{author}{Li, Z.} \&
  \bibinfo{author}{Farimani, A.~B.}
\newblock \bibinfo{title}{A physics-informed diffusion model for high-fidelity
  flow field reconstruction}.
\newblock \emph{\bibinfo{journal}{Journal of Computational Physics}}
  \textbf{\bibinfo{volume}{478}}, \bibinfo{pages}{111972}
  (\bibinfo{year}{2023}).

\bibitem{buzzicotti2023data}
\bibinfo{author}{Buzzicotti, M.}
\newblock \bibinfo{title}{Data reconstruction for complex flows using ai:
  recent progress, obstacles, and perspectives}.
\newblock \emph{\bibinfo{journal}{Europhysics Letters}}
  (\bibinfo{year}{2023}).

\bibitem{granero2024neural}
\bibinfo{author}{Granero-Belinchon, C.}
\newblock \bibinfo{title}{Neural network based generation of a 1-dimensional
  stochastic field with turbulent velocity statistics}.
\newblock \emph{\bibinfo{journal}{Physica D: Nonlinear Phenomena}}
  \textbf{\bibinfo{volume}{458}}, \bibinfo{pages}{133997}
  (\bibinfo{year}{2024}).

\bibitem{nichol2021improved}
\bibinfo{author}{Nichol, A.~Q.} \& \bibinfo{author}{Dhariwal, P.}
\newblock \bibinfo{title}{Improved denoising diffusion probabilistic models}.
\newblock In \emph{\bibinfo{booktitle}{International Conference on Machine
  Learning}}, \bibinfo{pages}{8162--8171} (\bibinfo{organization}{PMLR},
  \bibinfo{year}{2021}).

\bibitem{chevillard2003lagrangian}
\bibinfo{author}{Chevillard, L.} \emph{et~al.}
\newblock \bibinfo{title}{Lagrangian velocity statistics in turbulent flows:
  Effects of dissipation}.
\newblock \emph{\bibinfo{journal}{Physical review letters}}
  \textbf{\bibinfo{volume}{91}}, \bibinfo{pages}{214502}
  (\bibinfo{year}{2003}).

\bibitem{biferale2004multifractal}
\bibinfo{author}{Biferale, L.} \emph{et~al.}
\newblock \bibinfo{title}{Multifractal statistics of lagrangian velocity and
  acceleration in turbulence}.
\newblock \emph{\bibinfo{journal}{Physical review letters}}
  \textbf{\bibinfo{volume}{93}}, \bibinfo{pages}{064502}
  (\bibinfo{year}{2004}).

\bibitem{sohl2015deep}
\bibinfo{author}{Sohl-Dickstein, J.}, \bibinfo{author}{Weiss, E.},
  \bibinfo{author}{Maheswaranathan, N.} \& \bibinfo{author}{Ganguli, S.}
\newblock \bibinfo{title}{Deep unsupervised learning using nonequilibrium
  thermodynamics}.
\newblock In \emph{\bibinfo{booktitle}{International Conference on Machine
  Learning}}, \bibinfo{pages}{2256--2265} (\bibinfo{organization}{PMLR},
  \bibinfo{year}{2015}).

\bibitem{burda2015accurate}
\bibinfo{author}{Burda, Y.}, \bibinfo{author}{Grosse, R.} \&
  \bibinfo{author}{Salakhutdinov, R.}
\newblock \bibinfo{title}{Accurate and conservative estimates of mrf
  log-likelihood using reverse annealing}.
\newblock In \emph{\bibinfo{booktitle}{Artificial Intelligence and
  Statistics}}, \bibinfo{pages}{102--110} (\bibinfo{organization}{PMLR},
  \bibinfo{year}{2015}).

\bibitem{mordant2002long}
\bibinfo{author}{Mordant, N.}, \bibinfo{author}{Delour, J.},
  \bibinfo{author}{L{\'e}veque, E.}, \bibinfo{author}{Arn{\'e}odo, A.} \&
  \bibinfo{author}{Pinton, J.-F.}
\newblock \bibinfo{title}{Long time correlations in lagrangian dynamics: a key
  to intermittency in turbulence}.
\newblock \emph{\bibinfo{journal}{Physical review letters}}
  \textbf{\bibinfo{volume}{89}}, \bibinfo{pages}{254502}
  (\bibinfo{year}{2002}).

\bibitem{angriman2022multitime}
\bibinfo{author}{Angriman, S.}, \bibinfo{author}{Mininni, P.~D.} \&
  \bibinfo{author}{Cobelli, P.~J.}
\newblock \bibinfo{title}{Multitime structure functions and the lagrangian
  scaling of turbulence}.
\newblock \emph{\bibinfo{journal}{Physical Review Fluids}}
  \textbf{\bibinfo{volume}{7}}, \bibinfo{pages}{064603} (\bibinfo{year}{2022}).

\bibitem{mitra2004varieties}
\bibinfo{author}{Mitra, D.} \& \bibinfo{author}{Pandit, R.}
\newblock \bibinfo{title}{Varieties of dynamic multiscaling in fluid
  turbulence}.
\newblock \emph{\bibinfo{journal}{Physical review letters}}
  \textbf{\bibinfo{volume}{93}}, \bibinfo{pages}{024501}
  (\bibinfo{year}{2004}).

\bibitem{l1997temporal}
\bibinfo{author}{L'vov, V.~S.}, \bibinfo{author}{Podivilov, E.} \&
  \bibinfo{author}{Procaccia, I.}
\newblock \bibinfo{title}{Temporal multiscaling in hydrodynamic turbulence}.
\newblock \emph{\bibinfo{journal}{Physical Review E}}
  \textbf{\bibinfo{volume}{55}}, \bibinfo{pages}{7030} (\bibinfo{year}{1997}).

\bibitem{borgas1993multifractal}
\bibinfo{author}{Borgas, M.}
\newblock \bibinfo{title}{The multifractal lagrangian nature of turbulence}.
\newblock \emph{\bibinfo{journal}{Philosophical Transactions of the Royal
  Society of London. Series A: Physical and Engineering Sciences}}
  \textbf{\bibinfo{volume}{342}}, \bibinfo{pages}{379--411}
  (\bibinfo{year}{1993}).

\bibitem{nelkin1990multifractal}
\bibinfo{author}{Nelkin, M.}
\newblock \bibinfo{title}{Multifractal scaling of velocity derivatives in
  turbulence}.
\newblock \emph{\bibinfo{journal}{Physical Review A}}
  \textbf{\bibinfo{volume}{42}}, \bibinfo{pages}{7226} (\bibinfo{year}{1990}).

\bibitem{paladin1987degrees}
\bibinfo{author}{Paladin, G.} \& \bibinfo{author}{Vulpiani, A.}
\newblock \bibinfo{title}{Degrees of freedom of turbulence}.
\newblock \emph{\bibinfo{journal}{Physical Review A}}
  \textbf{\bibinfo{volume}{35}}, \bibinfo{pages}{1971} (\bibinfo{year}{1987}).

\bibitem{meneveau1996transition}
\bibinfo{author}{Meneveau, C.}
\newblock \bibinfo{title}{Transition between viscous and inertial-range scaling
  of turbulence structure functions}.
\newblock \emph{\bibinfo{journal}{Physical Review E}}
  \textbf{\bibinfo{volume}{54}}, \bibinfo{pages}{3657} (\bibinfo{year}{1996}).

\bibitem{benzi1993random}
\bibinfo{author}{Benzi, R.} \emph{et~al.}
\newblock \bibinfo{title}{A random process for the construction of multiaffine
  fields}.
\newblock \emph{\bibinfo{journal}{Physica D: Nonlinear Phenomena}}
  \textbf{\bibinfo{volume}{65}}, \bibinfo{pages}{352--358}
  (\bibinfo{year}{1993}).

\bibitem{guth2022wavelet}
\bibinfo{author}{Guth, F.}, \bibinfo{author}{Coste, S.},
  \bibinfo{author}{De~Bortoli, V.} \& \bibinfo{author}{Mallat, S.}
\newblock \bibinfo{title}{Wavelet score-based generative modeling}.
\newblock \emph{\bibinfo{journal}{Advances in Neural Information Processing
  Systems}} \textbf{\bibinfo{volume}{35}}, \bibinfo{pages}{478--491}
  (\bibinfo{year}{2022}).

\bibitem{salazar2009two}
\bibinfo{author}{Salazar, J.~P.} \& \bibinfo{author}{Collins, L.~R.}
\newblock \bibinfo{title}{Two-particle dispersion in isotropic turbulent
  flows}.
\newblock \emph{\bibinfo{journal}{Annual review of fluid mechanics}}
  \textbf{\bibinfo{volume}{41}}, \bibinfo{pages}{405--432}
  (\bibinfo{year}{2009}).

\bibitem{scatamacchia2012extreme}
\bibinfo{author}{Scatamacchia, R.}, \bibinfo{author}{Biferale, L.} \&
  \bibinfo{author}{Toschi, F.}
\newblock \bibinfo{title}{Extreme events in the dispersions of two neighboring
  particles under the influence of fluid turbulence}.
\newblock \emph{\bibinfo{journal}{Physical review letters}}
  \textbf{\bibinfo{volume}{109}}, \bibinfo{pages}{144501}
  (\bibinfo{year}{2012}).

\bibitem{biferale2005multiparticle}
\bibinfo{author}{Biferale, L.} \emph{et~al.}
\newblock \bibinfo{title}{Multiparticle dispersion in fully developed
  turbulence}.
\newblock \emph{\bibinfo{journal}{Physics of Fluids}}
  \textbf{\bibinfo{volume}{17}}, \bibinfo{pages}{111701}
  (\bibinfo{year}{2005}).

\bibitem{xu2011pirouette}
\bibinfo{author}{Xu, H.}, \bibinfo{author}{Pumir, A.} \&
  \bibinfo{author}{Bodenschatz, E.}
\newblock \bibinfo{title}{The pirouette effect in turbulent flows}.
\newblock \emph{\bibinfo{journal}{Nature Physics}}
  \textbf{\bibinfo{volume}{7}}, \bibinfo{pages}{709--712}
  (\bibinfo{year}{2011}).

\bibitem{roemmich2019future}
\bibinfo{author}{Roemmich, D.} \emph{et~al.}
\newblock \bibinfo{title}{On the future of argo: A global, full-depth,
  multi-disciplinary array}.
\newblock \emph{\bibinfo{journal}{Frontiers in Marine Science}}
  \textbf{\bibinfo{volume}{6}}, \bibinfo{pages}{439} (\bibinfo{year}{2019}).

\bibitem{essink2022characterizing}
\bibinfo{author}{Essink, S.}, \bibinfo{author}{Hormann, V.},
  \bibinfo{author}{Centurioni, L.~R.} \& \bibinfo{author}{Mahadevan, A.}
\newblock \bibinfo{title}{On characterizing ocean kinematics from surface
  drifters}.
\newblock \emph{\bibinfo{journal}{Journal of Atmospheric and Oceanic
  Technology}} \textbf{\bibinfo{volume}{39}}, \bibinfo{pages}{1183--1198}
  (\bibinfo{year}{2022}).

\bibitem{TURB-Lagr}
\bibinfo{author}{Biferale, L.}, \bibinfo{author}{Buzzicotti, M.},
  \bibinfo{author}{Bonaccorso, F.} \& \bibinfo{author}{Calascibetta, C.}
\newblock \bibinfo{title}{Turb-lagr. a database of 3d lagrangian trajectories
  in homogeneous and isotropic turbulence}.
\newblock \emph{\bibinfo{journal}{arXiv:2303.08662}}  (\bibinfo{year}{2023}).

\bibitem{calascibetta2023optimal}
\bibinfo{author}{Calascibetta, C.}, \bibinfo{author}{Biferale, L.},
  \bibinfo{author}{Borra, F.} \emph{et~al.}
\newblock \bibinfo{title}{Optimal tracking strategies in a turbulent flow}.
\newblock \emph{\bibinfo{journal}{Communications Physics}}
  \textbf{\bibinfo{volume}{6}}, \bibinfo{pages}{256} (\bibinfo{year}{2023}).

\bibitem{van2012efficiency}
\bibinfo{author}{Van~Hinsberg, M.}, \bibinfo{author}{Thije~Boonkkamp, J.},
  \bibinfo{author}{Toschi, F.} \& \bibinfo{author}{Clercx, H.}
\newblock \bibinfo{title}{On the efficiency and accuracy of interpolation
  methods for spectral codes}.
\newblock \emph{\bibinfo{journal}{SIAM journal on scientific computing}}
  \textbf{\bibinfo{volume}{34}}, \bibinfo{pages}{B479--B498}
  (\bibinfo{year}{2012}).

\bibitem{ronneberger2015u}
\bibinfo{author}{Ronneberger, O.}, \bibinfo{author}{Fischer, P.} \&
  \bibinfo{author}{Brox, T.}
\newblock \bibinfo{title}{U-net: Convolutional networks for biomedical image
  segmentation}.
\newblock In \emph{\bibinfo{booktitle}{Medical Image Computing and
  Computer-Assisted Intervention--MICCAI 2015: 18th International Conference,
  Munich, Germany, October 5-9, 2015, Proceedings, Part III 18}},
  \bibinfo{pages}{234--241} (\bibinfo{organization}{Springer},
  \bibinfo{year}{2015}).

\bibitem{vaswani2017attention}
\bibinfo{author}{Vaswani, A.} \emph{et~al.}
\newblock \bibinfo{title}{Attention is all you need}.
\newblock \emph{\bibinfo{journal}{Advances in neural information processing
  systems}} \textbf{\bibinfo{volume}{30}} (\bibinfo{year}{2017}).

\bibitem{song2021denoising}
\bibinfo{author}{Song, J.}, \bibinfo{author}{Meng, C.} \&
  \bibinfo{author}{Ermon, S.}
\newblock \bibinfo{title}{Denoising diffusion implicit models}.
\newblock In \emph{\bibinfo{booktitle}{International Conference on Learning
  Representations}} (\bibinfo{year}{2021}).
\newblock \urlprefix\url{https://openreview.net/forum?id=St1giarCHLP}.

\bibitem{lu2022dpm}
\bibinfo{author}{Lu, C.} \emph{et~al.}
\newblock \bibinfo{title}{Dpm-solver: A fast ode solver for diffusion
  probabilistic model sampling in around 10 steps}.
\newblock \emph{\bibinfo{journal}{Advances in Neural Information Processing
  Systems}} \textbf{\bibinfo{volume}{35}}, \bibinfo{pages}{5775--5787}
  (\bibinfo{year}{2022}).

\bibitem{li2024data}
\bibinfo{author}{Li, T.}, \bibinfo{author}{Biferale, L.},
  \bibinfo{author}{Bonaccorso, F.}, \bibinfo{author}{Scarpolini, M.~A.} \&
  \bibinfo{author}{Buzzicotti, M.}
\newblock \bibinfo{title}{Dataset for: Synthetic lagrangian turbulence by
  generative diffusion models}.
\newblock \bibinfo{howpublished}{Data set} (\bibinfo{year}{2024}).
\newblock \urlprefix\url{http://doi.org/10.15161/oar.it/143615}.

\bibitem{SmartTURB2024DiffusionLagr}
\bibinfo{title}{Smartturb/diffusion-lagr: stable} (\bibinfo{year}{2024}).
\newblock \urlprefix\url{https://doi.org/10.5281/zenodo.10563386}.

\bibitem{Li2024synthetic}
\bibinfo{author}{Li, T.}, \bibinfo{author}{Biferale, L.},
  \bibinfo{author}{Bonaccorso, F.}, \bibinfo{author}{Scarpolini, M.~A.} \&
  \bibinfo{author}{Buzzicotti, M.}
\newblock \bibinfo{title}{Supplementary code for: Synthetic lagrangian
  turbulence by generative diffusion models} (\bibinfo{year}{2024}).
\newblock \urlprefix\url{https://codeocean.com/capsule/0870187/tree/v1}.
\newblock \bibinfo{note}{CodeOcean}.

\bibitem{mordant2004experimental}
\bibinfo{author}{Mordant, N.}, \bibinfo{author}{L{\'e}v{\^e}que, E.} \&
  \bibinfo{author}{Pinton, J.-F.}
\newblock \bibinfo{title}{Experimental and numerical study of the lagrangian
  dynamics of high reynolds turbulence}.
\newblock \emph{\bibinfo{journal}{New Journal of Physics}}
  \textbf{\bibinfo{volume}{6}}, \bibinfo{pages}{116} (\bibinfo{year}{2004}).

\bibitem{homann2007lagrangian}
\bibinfo{author}{Homann, H.}, \bibinfo{author}{Grauer, R.},
  \bibinfo{author}{Busse, A.} \& \bibinfo{author}{M{\"u}ller, W.-C.}
\newblock \bibinfo{title}{Lagrangian statistics of navier--stokes and mhd
  turbulence}.
\newblock \emph{\bibinfo{journal}{Journal of Plasma Physics}}
  \textbf{\bibinfo{volume}{73}}, \bibinfo{pages}{821--830}
  (\bibinfo{year}{2007}).

\bibitem{biferale2005particle}
\bibinfo{author}{Biferale, L.}, \bibinfo{author}{Boffetta, G.},
  \bibinfo{author}{Celani, A.}, \bibinfo{author}{Lanotte, A.} \&
  \bibinfo{author}{Toschi, F.}
\newblock \bibinfo{title}{Particle trapping in three-dimensional fully
  developed turbulence}.
\newblock \emph{\bibinfo{journal}{Physics of Fluids}}
  \textbf{\bibinfo{volume}{17}}, \bibinfo{pages}{021701}
  (\bibinfo{year}{2005}).

\bibitem{fisher2008terascale}
\bibinfo{author}{Fisher, R.~T.} \emph{et~al.}
\newblock \bibinfo{title}{Terascale turbulence computation using the flash3
  application framework on the ibm blue gene/l system}.
\newblock \emph{\bibinfo{journal}{IBM Journal of Research and Development}}
  \textbf{\bibinfo{volume}{52}}, \bibinfo{pages}{127--136}
  (\bibinfo{year}{2008}).

\bibitem{yeung2006reynolds}
\bibinfo{author}{Yeung, P.}, \bibinfo{author}{Pope, S.~B.} \&
  \bibinfo{author}{Sawford, B.~L.}
\newblock \bibinfo{title}{Reynolds number dependence of lagrangian statistics
  in large numerical simulations of isotropic turbulence}.
\newblock \emph{\bibinfo{journal}{Journal of Turbulence}} \bibinfo{pages}{N58}
  (\bibinfo{year}{2006}).

\bibitem{xu2006high}
\bibinfo{author}{Xu, H.}, \bibinfo{author}{Bourgoin, M.},
  \bibinfo{author}{Ouellette, N.~T.}, \bibinfo{author}{Bodenschatz, E.}
  \emph{et~al.}
\newblock \bibinfo{title}{High order lagrangian velocity statistics in
  turbulence}.
\newblock \emph{\bibinfo{journal}{Physical review letters}}
  \textbf{\bibinfo{volume}{96}}, \bibinfo{pages}{024503}
  (\bibinfo{year}{2006}).

\bibitem{berg2006backwards}
\bibinfo{author}{Berg, J.}, \bibinfo{author}{L{\"u}thi, B.},
  \bibinfo{author}{Mann, J.} \& \bibinfo{author}{Ott, S.}
\newblock \bibinfo{title}{Backwards and forwards relative dispersion in
  turbulent flow: an experimental investigation}.
\newblock \emph{\bibinfo{journal}{Physical Review E}}
  \textbf{\bibinfo{volume}{74}}, \bibinfo{pages}{016304}
  (\bibinfo{year}{2006}).

\end{thebibliography}

\end{document}